\documentstyle[aps,epsfig]{revtex}
\begin{document}
\draft
{\Large \bf Effective interactions between electric double-layers}\\\\
  {{\large Jean-Pierre Hansen}\\Department of Chemistry, Cambridge
  University, Lensfield Road, Cambridge CB2 1EW, United Kingdom;
  e-mail: jph32@cus.cam.ac.uk\\\\
  {\large Hartmut L\"owen}\\Institut f\"ur Theoretische Physik II,
  Heinrich-Heine-Universit\"at, Universit\"atsstra\ss e 1,\\ D-40225
  D\"usseldorf, Germany; e-mail : hlowen@thphy.uni-duesseldorf.de}
{\large Shortened Title}: Electric double-layer interactions\\\\
{\large Abstract}\\
The present review summarizes and assesses recent 
theoretical and experimental advances, with special emphasis 
on the effective interaction
between charge-stabilized colloids, in the bulk or in 
confined geometries, and on the
ambiguities of defining an effective charge of the colloidal 
particles. Some consideration is given to the often neglected 
discrete solvent effects.\\\\
\newpage
\section{Introduction}
Electric double layers form spontaneously whenever surfaces carrying ionizable groups are 
suspended in a polar solvent, most frequently water.  The high
dielectric constant $\epsilon$ of the latter 
favours the dissociation of the functional surface groups, so that the surface acquires a net charge per 
unit area, $\sigma$. Oppositely charged counterions are released into the solvent, which generally also 
contains a finite concentration of microscopic anions and cations (microions), providing thus 
additional coions  and counterions.  The electric double-layer results from the 
build-up of a charge density (or "cloud") of opposite sign to that of the surface charge, which 
tends to screen the electrostatic potential due to the latter.  The width of the double-layer, which is a 
measure of its capacity, is determined by the competition between the thermal motion of the 
microions, which tends to spread out, or homogenize their distribution, in order to increase their 
entropy, and the electrostatic interactions, which attract the counterions towards the surface, while 
repelling the coions.
\\\\
The present review deals with electric double-layers around {\it mesoscopic} charged particles, which will 
be referred to as polyions (frequently also called macroions), and more specifically with the effective 
interactions between electric double-layers associated with different polyions.  Such double-layers 
are ubiquitous in many physical, geophysical, chemical and biological systems, including complex 
fluids, clay soils, polyelectrolytes (e.g. DNA) or cell membranes.  The main focus of the review will 
be on colloidal dispersions of spherical, rod-like or lamellar polyions, or micelles resulting from the 
self-assembly of ionic surfactants \cite{1}.  Advanced experimental techniques, like surface force 
machines or video microscopy in combination with optical tweezers, allow a direct measurement of 
the forces acting between charged surfaces or polyions.  Such direct measurements contrast with 
indirect determinations of effective forces via small angle X-ray or neutron scattering measurements 
of static structure factors \cite{2}.  For the theoretician, the main challenge is the highly asymmetric 
nature of dispersions involving mesoscopic polyions, and microscopic solvent molecules and ions.  
Clearly, at least three widely different length scales are involved, namely the characteristic size 
(radius) $a$  of the polyions, a typical width of the electric double-layer, of the order of the Debye 
screening length $\lambda_{D}$, and a typical correlation length $l$ of the solvent, of the order of a 
few molecular diameters.  Under most physical conditions the double
inequality $a\gg\lambda_{D}\gg l$  
holds, so that a coarse-grained statistical description of the suspension is clearly warranted.  This 
strategy may be cast in the unifying framework of the density functional theory (DFT) of 
inhomogeneous fluids \cite{3}, as explained in the following  section.  A number of key issues, which are 
presently the object of intense experimental and theoretical scrutiny, will be addressed in subsequent 
sections of this review.  They include the following questions and topics :
\begin{itemize}
\item	The limitations of the standard Poisson-Boltzmann or mean-field theory of electric double-
layers and the importance of spatial correlations between microions.  A particularly important 
question relates to the possibility of an attractive component of the effective interaction 
between polyions, induced by microion correlations.
\item	The notion of effective polyion charge (or charge renormalization), as controlled by 
dissociation equilibria and counterion adsorption (or "condensation"), and of charge 
regulation.
\item The relation between effective interactions between polyions, and the phase  behaviour of 
their dispersions.
\item	The effect of confinement on the effective interactions between colloids, and more generally 
the influence of electrostatic boundary conditions at interfaces.
\item	The introduction of solvent granularity into the statistical description of double-layers.
\end{itemize}
This review will emphasise the more recent experimental and theoretical developments in this very 
active field, covering mostly work published during the past decade.  Earlier work is adequately 
covered in some previous reviews of the subject \cite{4,5,6,7}.  Electrostatic CGS units will be used 
throughout.
\section{Multi-component versus effective one-component description}
Consider a suspension of $N$ polyions, of radius $a$ and charge $Z$, in a polar solvent with co and 
counterions, of radii $a_{\pm}$  and charges $z_{\pm}e$  ; the radius
of the solvent molecules is comparable to the  
microion radii $a_{\pm}$ , i.e. of the order of $0.1$ to 0.2 nm. In
micellar solutions, 1 nm $<a<10$ nm, and $10<|Z|<100$, while in most
colloidal dispersions, 10 nm $<a<10^{3}$ nm, and $10^{2}<|Z|<10^{4}$.   
A statistical description of the highly asymmetric multi-component system is a considerable 
challenge, which requires some degree of coarse-graining.  In most theoretical investigations of 
micellar or colloidal systems, the solvent is regarded as a mere continuum, characterized by its 
macroscopic dielectric constant $\epsilon$.  This amounts to a "primitive model" (PM) level of description, 
where polyions and microions are assumed to be charged hard spheres (non-spherical polyions will 
be considered later in this review) interacting via the Coulomb
potential $\sim e^{2}/\epsilon r$, outside the excluded 
volume range of inter-particle distances $r$.  It is implicitly
assumed that the mesoscopic polyions have  
the same macroscopic dielectric constant $\epsilon$ as the solvent,
thus avoiding complications due to  
dielectric discontinuities (e.g. image charges).  
\\\\
Even within the PM, the theoretician is still faced with the
polyion-microion asymmetry.  In micellar  
solutions, where the size and charge ratios $a/a_{\pm}$ and
$|Z|/|z_{\pm}|$ are roughly  a factor of 10, the asymmetry  
may still be handled within the multicomponent PM level of description, using the theoretical 
techniques of the theory of classical fluids \cite{8}.  In particular, the
partial pair distribution functions $g_{\alpha\beta}(r)$ (where
$\alpha$ and $\beta$ are
species indices) may be calculated from the usual fluid  
integral equations, including the hypernetted chain (HNC) \cite{9} and mean spherical approximations 
(MSA) \cite{10}, or their variants \cite{11} and hybrid combinations
\cite{12a,12b}.  Alternatively these correlation 
functions may be obtained from Monte Carlo (MC) or Molecular Dynamics (MD) simulations of the 
asymmetric PM.  The multicomponent PM point of view is, in practice, limited to micellar systems 
with $|Z|<10^{2}$, because integral equation closures are increasingly unreliable for asymmetric 
systems, and their numerical solution tends to become unstable.  Similarly simulations become more 
and more difficult, and are at present limited to $|Z|\le 60$ \cite{13}.    The multi-component PM 
description becomes untractable in the much more asymmetric colloidal range, which requires a 
coarse-grained description, based on effective interactions between polyions, to be discussed below.
\\\\
Important simplifications occur within the PM in a number of limiting situations.

\begin{enumerate}
\item[a)]	In the limit of infinitely low concentration $n$ (number per unit volume) of polyions, only a 
pair of these, within a solution of microions, need to be considered.  The effective interaction 
between the two polyions (p) is then exactly given by the potential of
mean force, $v_{\rm pp}(r) = -k_{\rm B}T \log [g_{\rm pp}(r)]$ \cite{14}.  The multicomponent problem reduces to that of an electrolyte of 
microions in the external field due to two fixed polyions, which leads to considerable 
simplifications in simulations \cite{15} or the numerical solution of
integral equations.

\item[b)]	The problem may be further simplified by taking the limit
  $a\rightarrow \infty$ ; the initial 
dispersion then reduces to the much simpler system of an electrolyte confined between two 
parallel planes carrying a, generally uniform, surface charge $\sigma$.  The classic 
problem of two interacting planar double-layers goes back to the work of Gouy \cite{16}, and has 
been the object of considerable theoretical and experimental work, some of which will be 
discussed in a subsequent section.

\item[c)]	In the opposite limit of high concentration, each polyion is, at least temporarily, trapped in 
the "cage" formed by neighbouring polyions.  This regime is reasonably
described by a one-body model, where a single polyion is located at
the centre of a Wigner-Seitz cell, together  
with an inhomogeneous distribution of microions, such that the total charge inside the cell is 
zero.  For practical purposes the cell is chosen of a simple geometry  reflecting the shape of 
the polyion (e.g. a spherically symmetric cell around a spherical polyion), and the 
electrostatic boundary conditions are chosen such as to mimic the average effect of the 
surrounding polyions \cite{17,18}.  Despite the considerable simplifications which they imply, 
 cell models allow valuable insight into concentrated dispersions, and will be discussed 
further on.
\end{enumerate}

A unifying framework for the statistical description of interacting electric double-layers, or, more 
generally, of inhomogeneous fluids and interfaces, is provided by density functional theory (DFT) 
\cite{3}.
\subsection{DFT and effective interactions}
In view of the considerable polyion-microion asymmetry, it seems natural to combine a discrete 
representation of the former with a field description of the latter.
Let $\{{\mathbf{R}}_{i}\}(1\le i \le N)$  denote the 
positions of the $N$ polyions, assumed here to be spherical, and let
$V^{{\rm dir}}_{N}\{{\mathbf{R}}_{i}\}$   be their direct 
interaction energy for a given configuration; $V^{{\rm dir}}_{N}$ is, to a good approximation, pairwise 
additive, with a pair potential $v_{\rm pp}(R)$ including a short-range excluded volume repulsion, the long-
range Coulomb repulsion $Z^{2}e^{2}/\epsilon R$  and a van der Waals attraction (dispersion force), of 
intensity characterized by a Hamaker constant \cite{1}.  The inhomogenous distributions of co- and 
counterions, in the "external" field of the polyions, are characterized by the local densities (or 
concentrations) $\rho_{\alpha}({\mathbf r})(\alpha=+,-)$.  The
equilibrium densities satisfy the variational principle \cite{3} :
\begin{eqnarray}
\frac{\delta\Omega\left [\rho^{*}_{+}(\mathbf r),\rho^{*}_{-}(\mathbf
    r) \right]}{\delta\rho^{*}_{\alpha}(\mathbf r)}{\Bigg |}_{
    \rho^{*}_{\alpha}=\rho_{\alpha}}=0;\;\;\;\;\;\;\alpha=+,-
\label{1}
\end{eqnarray}

where $\Omega$ is the grand potential functional of the trial densities  , given by :
\begin{eqnarray}
 \Omega\left[\rho^{*}_{+},\rho^{*}_{-}\right]=F\left[\rho^{*}_{+},\rho^{*}_{-}\right]-\sum_{\alpha}\int\Phi_{\alpha}({\mathbf
 r})\rho^{*}_{\alpha}({\mathbf r}){\rm d}\mathbf r 
\label{2}				
\end{eqnarray}
In (\ref{2}), $F$ denotes the {\it intrinsic} free energy functional,
while $\Phi_{\alpha}({\mathbf r})=\mu_{\alpha}-\Phi_{\alpha}^{\rm
  ext}({\mathbf r})$, with $\mu_{\alpha}$, the 
chemical potential of species $\alpha$ ,and $\Phi_{\alpha}^{\rm
  ext} $, the external potential acting on 
ions of this species, equal to the sum of the interactions with the $N$ polyions :
\begin{eqnarray}
 	\Phi_{\alpha}^{\rm
  ext}({\mathbf r})=\sum_{i=1}^{N} v_{{\rm p}\alpha}({\mathbf r}-{\mathbf R_{i}})
\label{3} 							
\end{eqnarray}
Note that in view of (\ref{3}), the equilibrium density profiles
$\rho_{\alpha}(\mathbf r)$, and the resulting grand 
potential $\Omega$,
depend parametrically on the polyion configuration.  The chemical
potentials  $\mu_{\alpha}$   are either fixed at some reservoir value (semi-grand canonical ensemble), 
or determined a posteriori by the canonical ensemble constraints:
\begin{eqnarray}
 \frac{1}{V}\int_{V}\rho_{\alpha}({\mathbf r}){\rm d}{\mathbf} r=n_{\alpha}
 \label{4}	
\end{eqnarray}
where $V$ is the total volume of the dispersion and $n_{\alpha}$ is the mean (macroscopic) 
concentration of $\alpha$-ions.
\\\\
Explicit calculations of the density profiles require some approximation for the generally unknown  
free  energy  functional  $F$,  which  is  traditionally split  into
ideal  and  excess parts,        $F = F_{\rm id} + F_{\rm ex}$;
$F_{\rm id}$ is known exactly :
\begin{eqnarray}
F_{\rm id}\left[\rho_{+},\rho_{-}\right]=k_{\rm
  B}T\sum_{\alpha=+,-}\int_{V}\rho_{\alpha}({\mathbf
  r})\left[\log\left(\Lambda^{3}_{\alpha}\rho_{\alpha}({\mathbf
  r})-1\right )\right ]{\rm d}\mathbf r	 
\label{5}				
\end{eqnarray}
If $n_{\alpha}^{(0)}$ denotes the concentrations of some homogeneous reference state (i.e. in the 
absence of polyions), and $\rho_{\alpha}({\mathbf
r};\xi)=n_{\alpha}^{(0)}+\xi\Delta\rho_{\alpha}({\mathbf
r})\;\;(0\le\xi\le 1) $ denotes a continuous set of density profiles,
such that $\rho_{\alpha}({\mathbf
r};\xi=1)$
  leads back to $\rho_{\alpha}(\mathbf
r)$, the equilibrium density profiles in the presence of 
polyions, then $F_{\rm ex}$ is given by the formally exact expression:
\begin{eqnarray}
	F_{\rm ex}\left[\rho_{+},\rho_{-}\right] & = & F_{\rm
	ex}\left(n_{+}^{(0)},n_{-}^{(0)}\right)+\sum_{\alpha}\mu_{\alpha}^{\rm
	ex}\int\Delta\rho_{\alpha}({\mathbf r}){\rm d}{\mathbf r} \\ \nonumber
        & - & k_{\rm
	B}T\sum_{\alpha}\sum_{\beta}\int_{0}^{1}d\xi(\xi-1) \\ \nonumber
        & \times & \int d
	{\mathbf r}\int {\rm d}  {\mathbf r^{'}}\Delta \rho_{\alpha}({\mathbf
	r})c_{\alpha\beta}\left[\left\{\rho_{\alpha}(\xi) \right \};\mathbf
	r,{\mathbf r^{'}}\right ]\Delta\rho_{\beta}(\mathbf r^{'}) 
\label{6}	 	
\end{eqnarray}
$\mu_{\alpha}^{\rm ex}$ is the excess part of the chemical potential,
while the $c_{\alpha\beta}\left[\left\{\rho_{\alpha}(\xi)\right\}\right]$ are the set of 
direct correlation functions for the inhomogeneous electrolyte with
local densities $\rho_{\alpha}(\mathbf r;\xi) $.  
For Coulombic systems, these decay asymptotically as
$-v_{\alpha\beta}(r)/k_{\rm B}T=-z_{\alpha
}z_{\beta}l_{\rm B}/r $ (where $l_{\rm B}=e^{2}/\epsilon k_{\rm B}T$  is the 
Bjerrum length).  It proves convenient to subtract this from the
$c_{\alpha \beta}$, leaving the short-
ranged part $c_{\alpha \beta}^{\rm sr}$ .  The excess free energy function then splits into a mean-field Coulombic 
contribution, and a correlation part :
\begin{eqnarray}
F_{\rm ex}\left[\rho_{+},\rho_{-}\right] & = & F_{\rm Coul}+F_{\rm
  corr} \\ \nonumber & = & \frac{e^{2}}{2\epsilon}\int {\rm d}{\mathbf
  r}\int {\rm d}{\mathbf
  r^{'}}\frac{\rho({\mathbf r})\rho({\mathbf r}^{'})}{|{\mathbf r}-{\mathbf
  r^{'}}|}+F_{\rm corr}
\label{7}                					
\end{eqnarray}
where $\rho({{\mathbf r}})=z_{+}\rho_{+}({{\mathbf r}})+z_{-}\rho_{-}({\mathbf r}) $  is the microion charge density.  The correlation term is formally given by (\ref{6}), 
with $c_{\alpha\beta}^{\rm sr}$  replacing $c_{\alpha\beta}$ .

Starting from these exact expressions, there are basically two strategies to proceed with 
approximations.

\begin{enumerate}
\item[a)]	The most common strategy focusses on the density profiles, and the resulting free energies.  
$F_{corr}$ is approximated by a local density ansatz (LDA) or by some weighted density 
approximation (WDA) \cite{3,19} which generally requires the direct correlation functions in a 
homogeneous reference state as input.  Once the variational problem (\ref{1}) has been solved for 
the approximate free energy functional, the {\it effective} interaction energy between polyions is 
given by :
\begin{eqnarray}
V_{N}\left(\{\mathbf R_{i}\}\right)=V^{\rm dir}_{N}\left(\{\mathbf R_{i}\}\right)+\Omega\left(\{\mathbf R_{i}\}\right)
\label{8}
\end{eqnarray}	
					
	where the equilibrium grand potential accounts for their indirect interactions, induced by the 
double-layer-forming microions, which have effectively been traced out.  It is very important 
to realize that this contribution depends on the thermodynamic state of the dispersion, and is 
generally not pair-wise additive.

	If correlations are neglected altogether, $F_{\rm ex}$ reduces to the mean-field Coulombic part, and (\ref{1}) 
immediately leads to the equilibrium profiles 
\begin{eqnarray}
\rho_{\alpha}({\mathbf
r})=\zeta_{\alpha}\exp\left\{-\left[\Phi_{\alpha}^{\rm dir}({\mathbf
    r})+z_{\alpha}e\psi({\mathbf r})\right]/k_{\rm B}T \right\}
\label{9}	 						
\end{eqnarray}

	where $\zeta_{\alpha}$  is the fugacity of species $\alpha$  (equal to its reservoir 
concentration in the absence of correlations), $\Phi_{\alpha}^{\rm sr}$  	 is the 
short-range (excluded volume) part of the interaction of an ion $\alpha$  with the $N$ 
polyions, and $\psi (\mathbf r)$  is the total electrostatic potential
at $\mathbf r$, which satisfies Poisson's 
equation :
\begin{eqnarray}
\nabla^{2}\psi({\mathbf r})=-\frac{4\pi e}{\epsilon}\left[\rho^{\rm ext
    }({\mathbf r})+\rho ({\mathbf r})\right]	
\label{10}						
\end{eqnarray}

	with $\rho^{\rm ext
    }(\mathbf r)$  the "external" charge density carried by the $N$ polyions.  Eqs (\ref{9}) and 
(\ref{10}) constitute the multi-centre version of mean-field non-linear Poisson-Boltzmann (PB) 
theory.  Despite their apparent simplicity, numerical solution of the PB equations in the 
presence of $N$ polyions is a formidable task.  In practice, eq. (\ref{10}) is often solved in the 
domain outside the polyions, and the contribution of the latter to the electrostatics is treated 
as a boundary value problem.

\item[b)]	A second, more ambitious strategy is to seek
  information on both the density profiles, and the  
pair correlations between microions within the double-layers.  This may be achieved by 
relating the direct correlation functions, appearing in (\ref{6}), to
their functional inverses, the total  
correlation functions $h_{\alpha\beta}$ , via the Ornstein-Zernike relations for inhomogeneous 
fluids \cite{8}, supplemented by an approximate closure relation between the h and c functions.  
This strategy leads to coupled equations for the density profiles and the pair correlation 
functions, as in the widely used inhomogeneous hypernetted chain (IHNC) theory \cite{20}.  In 
view of its numerical complexity, this strategy can only be applied to very simple geometries.  
In practice it is limited to planar geometry and will be discussed in a subsequent section.

\end{enumerate}

\subsection{Cell model}
As stated earlier, cell models, involving a single polyion, prove useful to study concentrated 
dispersions, including colloidal crystals.  The paradigm is provided by the much studied case of a 
globular polyion of radius $a$, placed at the centre of a spherical cavity of radius $R$ determined by the 
polyion concentration, i.e. $R=(3/4\pi n)^{1/3}$, containing counterions and salt which ensure overall 
charge neutrality.  The density profiles $\rho_{+}(r)$ and $\rho_{-}(r)$ depend only on the distance $r$ from the 
centre (1d problem).  Gauss's law implies the boundary conditions (b.c.) that the electric field vanish 
on the surface of the cell, i.e. $d\psi/dr|_{r=R}=0$.  The field on the surface of the particle (i.e. at $r = a$) is 
determined by the surface charge density $\sigma$ .  Finally, due to the b.c. at $r = R$, the osmotic pressure $P$ 
of the microions is given by \cite{18} :
\begin{eqnarray}
P=k_{\rm B}T[\rho_{+}(R)+\rho_{-}(R)]
\label{11}	 								
\end{eqnarray}
PB theory, embodied in eqs. (\ref{9}) and (\ref{10}), reduces here to a spherically symmetric one-centre 
problem, giving rise to a simple second order non-linear differential equation for the local potential $\psi(r)$
 , which is easily solved numerically \cite{21,22}.  Improvements over the mean-field approximation 
are achieved by including correlations within the LDA or the WDA
\cite{22,23}.  The resulting osmotic 
pressures are consistently lower than those predicted by PB theory, by a factor of three or more for 
highly charged polyions.  This trend is confirmed by MC simulations of the microions in the cell, 
subjected to the electrostatic potential of the polyion \cite{22,23}.
\\\\
Cell model calculations have been extended to other geometries, e.g. rods or platelets in cylindrical 
cells.  They prove very useful in the determination of 
{\it effective} polyion charges, as discussed in the next section.
\section{Charge regulation and renormalization}
Most published work on electric double-layers is based on the assumption of constant charge on the 
polyions, e.g. in the form of a constant and uniform surface charge.  This is clearly an 
oversimplification, as would be the other extreme assumption of constant surface potential, for two 
reasons :  the strong coupling between electrostatics and chemical dissociation equilibrium at the 
surface ({\it charge regulation}), and the adsorption and strong physical binding of counterions to the 
surface, often referred to as counterion condensation, which leads to a reduction of the apparent 
polyion charge seen at larger distances ({\it charge renormalization}).
\subsection{Charge regulation}
The bare (or structural) surface charge of most polyions results from the dissociation of 
functional groups like sulphate or carboxylic groups, the number of which may be measured by 
titration \cite{24}.  Generally the ionization is only partial, and the dissociation equilibrium, governed by 
a law of mass action, depends on the local ionic environment, e.g. on local salt concentration or pH.  
Any variation of this environment, linked to the relative motion of neighbouring polyions, will 
lead to a fluctuation of the surface charge, but for given macroscopic conditions, one may define 
some average bare charge.  From a theoretical point of view, the coupling of the surface chemistry to 
the local inhomogeneities induced by electrostatics poses a difficult challenge.  Early attempts were 
based on a combination of the law of mass action with PB theory \cite{25}.  A more microscopic point of 
view is adopted in the "charge regulation primitive model" (CRPM) \cite{26} (see also Ref \cite{26a}), which 
adds a strong attraction of chemical origin to the short-range part
$v_{{\rm p}\alpha}^{sr}$  of the polyion-microion pair 
potential, such that :
\begin{eqnarray}
\exp\left\{-v_{{\rm p}\alpha}^{sr}(r)/k_{\rm B}T\right\} & = &
V_{\alpha}\delta(r-d_{\alpha}); \;\;\;\;\;\;\;
r<\frac{1}{2}(a+a_{\alpha}) \\ \nonumber & = & 1;
\;\;\;\;\;\;\;\;\;\;\;\;\;\;\;\;\;\;\;\;\;\;\;\;
r>\frac{1}{2}(a+a_{\alpha}) 
\label{12}	 				
\end{eqnarray}
This corresponds to an infinitely deep and narrow potential well localized on a sphere of radius $d_{\alpha}$  
from the centre of the polyion;  in practice $d_{\alpha}$ must be chosen to be smaller than the polyion radius 
$a$, to prevent the chemical binding of the same microion to two polyions.\\\\
The pair structure of the model has been analysed by diagrammatic expansion, and by numerical 
solutions of the HNC equation for small polyions corresponding to mineral oxide particles \cite{26}.  The 
charge regulation strongly affects the effective interaction between spherical particles, to be 
discussed later.
\subsection{Charge renormalization}
The bare (or structural) charge resulting from the dissociation equilibrium is frequently very large, 
typically $|Z|\ge10^{4}$  for polyion radii $a\ge 10^{2}$ nm.  To
describe electric double-layers near such highly  
charged polyions, the traditional phenomenological approach is to divide the counterions into two 
populations.  The first includes ions that are tightly bound (or adsorbed) to the surface by the strong 
electric field $E=4\pi\sigma/\epsilon$, thus forming a so-called Stern
layer of "condensed" counterions.  A rough  
estimate of the thickness $\Delta$  of a Stern layer is obtained by
balancing the electrostatic work $E\times ze\times \Delta $ 
  against the thermal energy $k_{\rm B}T$. The resulting
  $\Delta=\epsilon k_{\rm B}T/4\pi\sigma z e$ is of the order of a
  few \AA$\;$  in water at room temperature for typical surface charge densities and monovalent counterions;  
thus  $\Delta$  is comparable to the size of microions, so that the "condensate" may be expected to be 
roughly a monolayer.  This monolayer, of opposite sign to the surface charge, strongly compensates 
the latter, reducing the total polyion charge to an effective charge $Z^{*}$ significantly smaller than the 
bare charge $Z$.  The remaining counterions feel a much reduced "external" potential and form the 
"diffuse" part of the double-layer, which can often be treated within linearized PB theory (LPB).\\\\
The phenomenological approach is to consider $Z^{*}$ as a parameter adjusted to experimental data, e.g. 
polyion structure factors determined by light scattering \cite{2,27a,27b,27c}, but much recent theoretical 
effort has gone into determining the effective charge from first principles.  Most schemes are based 
on the observation that the asymptotic behaviour of the potential or density profiles is correctly 
described by the simple exponential screening predicted by LPB theory.  Charge renormalization 
should account both for non-linearities in the mean-field PB approach and for microion correlations.  
Early attempts focused on the single polyion problem within the cell
model \cite{21,22}.  Within LPB 
theory the electrostatic potential in a spherical cell is easily calculated to be of the form :
\begin{eqnarray}
\psi(r)=C+\frac{Ze}{\epsilon
  r}\left(Ae^{\kappa_{D}r}+Be^{-\kappa_{D}r} \right)
\label{13}
\end{eqnarray}
where $\kappa_{D}=1/\lambda_{D}$  is the inverse Debye screening length :
\begin{eqnarray}
	 \kappa_{D}=\left(4\pi l_{\rm
	 B}\sum_{\alpha}n_{\alpha}z_{\alpha}^{2}\right )^{1/2}
	\label{14}
\end{eqnarray}
 The integration constants $A$ and $B$ are determined by the b.c. at $r = a$ 
and $r = R$;  $C$ is conveniently chosen such that $\psi(r=R)=0$ .  The validity of the LPB  
approximation is expected to be reasonable on the cell surface, but to be very poor near the polyion 
surface if the latter is highly charged.  The bare 
charge $Z$ appearing in the LPB potential (\ref{13}) is then adjusted to a (lower) effective value $Z^{*}$ by 
requiring that the resulting microion charge density on the cell
surface, $\rho(R)$, match the 
corresponding density obtained from a numerical integration of the PB equation under the same b.c. 
\cite{21}.  In view of the latter, and of Poisson's equation, this means that the resulting LPB potential 
matches the non-linear PB potential at the surface up to the third derivative.  Note that a 
renormalization of $Z$ also implies a renormalization of $\kappa_{D}$  at low (or vanishing) concentration of 
added salt, due to the overall charge neutrality requirement.  In the absence of salt  (counterions only), the matching condition also implies equality of osmotic pressures.  Since LPB theory 
underestimates the charge density near the polyion surface, it overestimates $\rho(r=R)$ , compared to 
the PB charge profile, so that $Z^{*}$ is always lower than
$Z$. $Z^{*}\rightarrow Z$ for low polyion charge, then  
increases with $Z$, but {\it saturates} for high values of the bare charge, as a consequence of counterion 
condensation;  the degree of counterion condensation is not greatly affected by salt concentration 
\cite{21}.  The saturation value of $Z^{*}$ is roughly proportional to the polyion radius $a$.\\\\
The effects of microion correlations on the values of $Z^{*}$ have been investigated with LDA and WDA 
versions of DFT, and by MC simulations \cite{22,23}.  Correlations tend to
lead to a further decrease of 
$Z^{*}$ compared to $Z$, and even to a maximum value, followed by an actual decrease of $Z^{*}$ as a function 
of $Z$ \cite{22}.  If the effective charge is obtained by matching the LPB osmotic pressure to "exact" MC 
estimates, the resulting $Z^{*}$ turns out to depend only weakly on the ratio $a/R$, i.e. on the concentration 
of polyions.\\\\
Other semi-phenomenological criteria for determining the effective
charge are discussed in ref. \cite{28}.  
The Debye-H\"uckel-Bjerrum theory, that has proved very successful to describe the phase behaviour 
of PM electrolytes \cite{29} has been extended to charged colloids \cite{30}.  The Bjerrum pairs of simple 
electrolytes are replaced by clusters involving each a polyion and a variable number of adsorbed 
counterions, which are in chemical equilibrium with the free, non-adsorbed counterions, assuming a 
simple form for the internal partition function of a cluster.  The distribution of cluster sizes turns out 
to be sharply peaked, allowing a rather clear-cut definition of $Z^{*}$ \cite{30}.\\\\
A more fundamental approach to a definition of effective charges is based on a careful analysis of 
the asymptotic decay of density profiles and correlation functions.  This may be achieved within the 
"dressed ion" reformulation of linear response theory, i.e. of the polarization of an electrolyte by the 
average electrostatic potential \cite{31}.  Within the PM, the point charges of the microions are replaced 
by a local, spread-out charge distribution incorporating the short-range fraction of the polarization 
charge around each bare ion, which is thus replaced by a dressed ion.  The short-range fraction of the 
polarization charge is defined in terms of short-range contributions
$h_{\alpha\beta}^{\rm sr}$  to the pair correlation 
functions $h_{\alpha\beta}(r)=g_{\alpha\beta}(r)-1$ ;  the
$h_{\alpha\beta}^{\rm sr}$  are related to the short-range parts $c_{\alpha\beta}^{\rm sr}$  of
the direct  
correlation functions, introduced after eq. (\ref{6}), via coupled Ornstein-Zernike (OZ) relations \cite{32}.  
This leads to an exact, non-local generalization of the LPB equations, and a residue analysis of the 
corresponding Fourier transform of the potential $\psi_{i}(r)$  around an ion i leads to the following 
asymptotic decay for large $r$ \cite{31,32}:
\begin{eqnarray}
\psi_{i}(r)\sim\frac{z_{i}^{\ddag}e \exp({-\kappa r})}{\epsilon^{\ddag} r}
\label{15}
\end{eqnarray}
This is precisely of the LPB (or Debye-H\"uckel) form, but with
renormalized values $z_{i}^{\ddag},\kappa$ and $\epsilon^{\ddag}$  of 
the valence, inverse screening length and dielectric constant of the solution (as opposed to the pure 
solvent), which can all be expressed in terms of the
$h_{\alpha\beta}^{\rm sr}$ ;  the latter may be calculated by  
supplementing the OZ relations with some approximate closure, e.g. HNC.\\\\
In the weak-coupling limit of very low ionic bulk concentration,
$z^{\ddag}_{i}\rightarrow z_{i}$, $\kappa\rightarrow \kappa_{\rm D}$ and
$\epsilon^{\ddag}\rightarrow \epsilon$.  
Simple Stillinger-Lovett sum rule \cite{33} considerations show that proper
inclusion of the finite size $a_{\alpha}$   
of the ions leads to an enhancement of screening,
i.e. $\kappa>\kappa_{\rm D}$ \cite{34}.  For sufficiently strong coupling 
(i.e. at high ionic concentrations), the exponential decay (\ref{15}) goes over to a damped oscillatory 
behaviour characterized by a pair of {\it complex} conjugate inverse
decay lengths $\kappa$ and $\kappa^{*}$  \cite{34,35,36}.  
The location of the cross-over from monotonous to oscillatory decay depends on the approximate 
closure.\\\\
The "dressed ion" reformulation for the bulk PM may be extended to the case where one ionic 
species consists of polyions, and hence $z_{i}^{*}$  is to be
identified with the effective charge $Z^{*}$ of the latter  
\cite{32}.  A similar treatment has been applied to planar electric double layers, to determine effective 
surface charge densities  \cite{37}.  The latter tend to saturate for monovalent counterions, increasingly so 
as their concentration increases, while for concentrated solutions of divalent ions, the effective 
surface charge goes through a maximum before decreasing as the bare surface charge increases.  
This behaviour is reminiscent of that observed in a spherical cell
\cite{22,23}.\\\\
The effective polyion charge $Z^{*}$ is not easily accessible to experiment.  Indirect determinations are 
based on the assumption of the validity of some simple functional form of the effective pair 
interaction between polyions, generally a screened Coulomb (or DLVO) potential.  The 
thermodynamic properties and pair structure of a system of polyions interacting via this potential can 
be determined accurately from fluid integral equations or
simulations \cite{38a,38b}.  The effective 
charge $Z^{*}$ is then adjusted to provide the best theoretical fit to experimental data, e.g. the polyion 
pair structure as measured by scattering experiments\cite{2,27a,27b,27c}, or the freezing line as a function of salt 
concentration and bare particle charge \cite{39}.  The latter measurements confirm the saturation 
predicted by the PB cell model \cite{21,22}, but should be re-analysed since the underlying simulation 
data do not take into account an important "volume" contribution to the free energies, which strongly 
affects the phase diagram at low salt concentration \cite{40a,40b}.\\\\
Recent experiments, carried out at very low electrolyte concentration, on silica particles, with weakly 
dissociating silanol surface groups, and on latex particles, with strongly dissociating sulfonic acid 
groups, were able to measure simultaneously the bare polyion charges by conductometric titration, 
and the effective charges by conductivity measurements \cite{41a,41b}.  The bare charge was controlled by 
varying the amount of added NaOH.  Under these experimental conditions, $Z^{*}$ was not observed to 
saturate, but to follow very nearly a square root law $Z^{*}\sim \sqrt{Z}$  for both colloidal systems.
\section{Planar geometry}
The simplest, and most widely studied double-layer geometry is that of uniformly charged, infinite 
planes in an ionic solution \cite{16}.  These planes may represent stacks of thin charged lamellae, like 
clay platelets or rigid membranes, or correspond to the surfaces of spherical colloidal particles 
separated by a distance $h\ll a$,  so that the curvature of the facing surfaces may be neglected in first 
approximation.  In the former situation, one may consider the simplified model of a single charged 
plane within a slab of thickness h equal to the mean spacing between lamellae in the stack;  this 
would be the 1d equivalent of the cell model.  The two situations are
shown schematically in Figure 1;    
 $\sigma$ denotes the surface charge density, $\epsilon$  is the
 macroscopic dielectric constant of the solvent (within a PM
 representation)
and $\epsilon^{'}$  is the dielectric constant of the colloidal
particles.  When $\epsilon^{'}\neq \epsilon$, which is the rule rather than 
the exception, the dielectric discontinuity at $z = \pm h/2$  must be properly incorporated in the boundary 
condition, e.g. by the introduction of electrostatic image charges \cite{43}.  In most published theoretical 
work the assumption $\epsilon^{'}=\epsilon $ is made for the sake of simplicity.\\\\
Microion density profiles $\rho_{+}(z)$ and $\rho_{-}(z)$  depend only
on $z$, and the intrinsic free energy functional  
per unit area perpendicular to the $z$ axis is still given by eqs
(\ref{5})-(\ref{7}), with the volume integrals   
replaced by 1d integrals over the interval $[-\frac{h}{2},\frac{h}{2}]$ .  Fixed
charge (or Neumann) boundary conditions imply that the potential
satisfy : 
\begin{eqnarray}
	-\frac{d\psi}{dz}\bigg |_{z=\pm \frac{h}{2}} =\pm \frac{4\pi\sigma}{\epsilon}
\label{16}
\end{eqnarray}
The key physical quantity, which is in principle measurable using a surface force apparatus \cite{1}, is the 
force per unit area acting between two charged planes and their associated electric double-layers.  
This force per unit area is the net osmotic pressure (or disjoining
pressure) $\Delta P=P-P_{\rm bulk}$, where $P$ 
is the osmotic pressure exerted by the microions between the charged
plates, and $P_{\rm bulk}$ is the pressure 
of the electrolyte in the reservoir which fixes the chemical potentials of the microions.  $P$ may be 
calculated via the mechanical route, by averaging the local stress tensor, or by the thermodynamic 
route, by differentiation of the total free energy, or grand potential, with respect to $h$.  Both routes 
lead to a natural separation of the pressure $P$ into kinetic, electrostatic and collisional parts \cite{43} :
\begin{eqnarray}
 P=P_{\rm kin}+P_{\rm el}+P_{\rm coll}	 	 							
\label{17}
\end{eqnarray}	
where
\begin{eqnarray}	 		
P_{\rm kin} = k_{\rm B}T\sum_{\alpha}\rho_{\alpha}(z)							
\label{18}
\end{eqnarray}	
and explicit expressions for $P_{\rm el}$ and $P_{\rm coll}$ in terms of density profiles and inhomogeneous pair 
correlation functions are given in ref. \cite{43}.  Note that, while each of the terms in eq. (\ref{17}) depends on 
the coordinate $z$, their sum must be independent of z for mechanical equilibrium.  In practice, 
simplified expressions are obtained by calculating $P$ in mid-plane ($z = 0$ in Fig. 1) or at the surface 
of the charged planes.  In the latter case the expression for $P$ reduces to the simple contact theorem 
\cite{5} :
\begin{eqnarray}
 P=k_{\rm
 B}T\sum_{\alpha}\rho_{\alpha}\left(z=\pm(h/2-a_{\alpha})\right)-\frac{2\pi\sigma^{2}}{\epsilon}  
\label{19}
\end{eqnarray}
This expression is not very convenient for simulation purposes, since $P$ appears as a difference of 
two large numbers, the first of which involves the contact value of the counterion density 
profile, affected by large numerical uncertainties.  The calculation at midplane is then preferable, 
since the $\rho_{\alpha}$  are expected to be close to their bulk values there, but the average force between 
microions on both sides of this plane must be evaluated \cite{44}.
\subsection{Poisson-Boltzmann approximation}
The mean-field approximation, where correlations between microions, due to excluded volume and 
Coulomb interactions, are neglected, is well understood since the pioneering work of Gouy \cite{16};  
density profiles and forces are known analytically, or given by simple quadratures \cite{45}.  The 
pressure is given throughout by : 
\begin{eqnarray}
 P=k_{\rm
 B}T\sum_{\alpha}\rho_{\alpha}(z)-\frac{\epsilon}{8\pi}\left[E(z)\right]^{2}
\label{20}
\end{eqnarray}
where $E(z)=-d\psi(z)/dz$  is the mean local electric field.  This is most easily evaluated in the mid-
plane, where $E = 0$  by symmetry, i.e. $P=k_{\rm
  B}T\sum_{\alpha}\rho_{\alpha}(0)$ , showing that the force between
equally charged plates is always {\it repulsive} within PB theory.\\\\
In the limit of low surface charge {\it and} electrolyte concentration, PB theory may be linearized, and the 
disjoining pressure reduces to :
\begin{eqnarray}
	 \Delta P=2\sigma^{2}e^{-\kappa_{\rm D}h}	
\label{21}
\end{eqnarray}
A charge renormalization to account for non-linearities, very similar to the procedure within the cell 
model described earlier, leads to a much reduced effective surface charge $\sigma^{*}$, and to a saturation for 
large bare charges $\sigma$ \cite{46}.
\subsection{The effect of microion correlations}
While PB theory always predicts a purely repulsive interaction between planar electric double-layers, 
it was suggested by Oosawa as early as 1968 \cite{47} that the force might turn attractive if microion 
correlations were taken into account.  The first convincing evidence for double-layer attraction came 
from MC simulations which indicated large deviations from the predictions of PB theory in the 
presence of {\it divalent} counterions \cite{44}.  This early prediction of an attractive minimum in the 
effective force between planar double-layers at short separations h was confirmed by subsequent 
simulations in the presence of salt ($1 : 2$ and $2 : 2$ electrolytes)
\cite{43,48} and polyvalent counterions \cite{delville2}, by inhomogeneous HNC 
calculations \cite{43}, and an improved version based on the reference HNC (RHNC) closure \cite{49,50}, 
and by LDA and WDA versions of DFT \cite{51,52}.\\\\
The  reduction of the double-layer repulsion and eventual correlation-induced attraction may be 
understood by considering the contact theorem (\ref{19}), or the expression for the pressure in midplane.  
At contact the correlations keep the counterions in the first layer apart, thus limiting the piling up 
following from the uncorrelated mean field treatment, and reducing the contact value of the 
counterion density;  according to eq. (\ref{19}) this leads to a lowering of the pressure, which can even 
become negative (corresponding to an effective attraction), due to the negative Coulombic 
contribution.  The lowering of the counterion concentration immediately at contact leads to an 
enhanced attraction of the next layers of counterions to the surface, so that correlations enhance the 
counterion density in that region.  This enhancement in turn entails a depletion of the counterion 
density in the midplane, while electrostatic correlations make a
negative contribution ($P_{\rm el}$ in eq. (\ref{17})) 
which generally dominates the smaller, positive collisional part.  Overall these various effects lead to 
a much reduced $\Delta P$  compared to the repulsive PB interaction.  The attraction observed for very short 
separations may be explained by a 2d lattice-like structuring of the adsorbed counterions, when the 
counterion patterns on opposite surfaces are shifted relative to each other to minimize the 
electrostatic interactions \cite{53}.\\\\
The reason why an attraction is in general observed only with divalent
(e.g. ${\rm Ca^{++}}$) counterions is that 
the entropic cost associated with their "condensation" near the surface is smaller than for monovalent 
ions, since only half as many counterions are needed to provide the same electrostatic shielding.  
Another related aspect of microion correlations is the phenomenon of charge reversal (or inversion).  
The apparent charge density of the surface placed at $z=-h/2$, seen at the abscissa $z$ is \cite{54} :
\begin{eqnarray}
\sigma^{*}(z)=\sigma+e\int_{-\frac{h}{2}}^{z}\sum_{\alpha}z_{\alpha}\rho_{\alpha}(z^{'})dz^{'}	 							
\label{22}
\end{eqnarray}
Within PB theory  $\sigma^{*}$ never changes sign, but in the presence of correlations this may occur at a 
critical z, beyond which  $\sigma^{*}$  is of opposite sign to
$\sigma$ , due to an over-compensation of the bare surface  
charge by the counterions.  Thus a test-charge will be attracted, rather than repelled, by a plane 
carrying a surface charge $\sigma$  of the same sign, beyond a critical separation.  Charge inversion can 
only occur when the microion density profiles have a non-monotonic behaviour.  The over-
compensation of the surface charges of two parallel plates is enhanced when the counterions are 
linked together to form polyelectrolyte chains, due once more to the reduction of entropic repulsion 
\cite{54,55}.\\\\
Charge inversion and effective attraction between platelets may even be observed in the case of 
monovalent counterions, provided the coions are larger than the counterions \cite{50}.\\\\
Attractive forces between charged surfaces immersed in aqueous
solutions of ${\rm CaCl_{2}}$ and ${\rm Ca(NO_{3})_{2}}$  
have been measured directly with a surface force apparatus \cite{56}.  The existence of correlation-induced 
attraction between charged surfaces of equal sign has important consequences on colloid stability, 
and provides a plausible explanation for the common observation that the addition of multivalent 
ions to solutions or suspensions of polyions often leads to precipitation.
\section{Spherical polyions}
Aqueous solutions and dispersions of spherical or quasi-spherical polyions, including globular 
proteins and charge-stabilized colloidal particles, like silica mineral particles or polymer latex 
spheres, have been thoroughly studied experimentally and theoretically for many decades.  Much of 
the interest stems from the realization that such colloidal systems exhibit a phase behaviour 
reminiscent of that of simple molecular systems, albeit on very different scales.  A particularly 
attractive feature of colloidal dispersions is that the effective interactions between particles may be 
{\it tuned} by the experimentalist, e.g. by varying the concentration of added electrolyte, thus providing 
an additional handle on phase behaviour.  The beautiful early microscopy observations by Hachisu et 
al. \cite{57}  unambiguously showed the existence, at low volume fractions, of colloidal crystals and 
crystalline alloys, which Bragg-reflect visible light and are, in particular, responsible for the 
iridescence of opals.  \\\\
The coexistence of ordered crystalline and disordered fluid phases has been monitored by very 
careful experiments using confocal laser scanning microscopy, which allows the direct observations 
of ordered and disordered configurations, and ultra-small-angle X-ray scattering (USAXS), which 
provides statistically averaged information on ordering of colloids via the static structure factor $S(k)$ 
\cite{41a,41b}.  These measurements provide three-dimensional phase diagrams by varying the colloid 
concentration $n$ or packing fraction $\eta=4\pi n a^{3}/3$, the
effective colloid charge $Z^{*}$ (as determined by  
conductivity measurements), and the monovalent salt concentration $n_{s}$, equal to the concentration $n_{-}$ 
of coions, typically in the range $10^{-6}-10^{-5}$M.  The observed
phase diagrams exhibit a striking {\it re-entrant} behaviour, e.g. for
fixed $n$ and $n_{s}$, the colloidal dispersion crystallizes into an
ordered BCC  
lattice upon increasing $Z^{*}$, as one might expect, but remelts into a disordered fluid phase upon 
further increase of $Z^{*}$, .  This re-entrant behaviour may be qualitatively understood in terms of an 
increase of the total ionic strength, linked to an increase of the counterion concentration $n_{+}$ with $Z^{*}$, 
which enhances the screening power of the microions and hence reduces the range of the screened 
Coulomb repulsion between polyions.  However, the standard DLVO representation of the latter \cite{58} 
fails to provide a quantitative explanation of the experimental data \cite{59}.\\\\
Another striking feature of low concentration dispersions
$(\eta<0.05)$ , reported by the Kyoto School,  
is the appearance of strongly inhomogeneous patterns in highly deionized samples.  USAXS 
measurements and confocal laser scanning microscopy show clear evidence of regions of relatively 
high colloid concentration, which may be crystalline \cite{60}, amorphous \cite{61} or fluid \cite{62} in character, 
coexisting with regions of extremely low concentration, or {\it voids}.  Apart from direct visual 
observation, the existence of voids, and the fraction of the total volume occupied by them, may be 
inferred from a comparison between the mean inter-particle spacing,
$d_{n}\sim n^{-1/3}$, calculated under the 
assumption of a homogeneous dispersion, and the spacing $d_{x}$
deduced from the position of the main  
X-ray diffraction peak in the structure factor $S(k)$.  The observation that $d_{x}$ is significantly less than $d$ 
points to the existence of voids occupying a fraction $f = 1 -(d_{x}/d)^{3}$  of the total sample volume.  The 
considerable literature on the subject is summarized in ref \cite{63}.  The existence of voids, and the 
related, but controversial \cite{64}, observation of a complete separation between a high concentration 
colloidal "liquid"    phase, and a much more dilute "gas" phase \cite{65}, are claimed to be evidence for a 
long range {\it attractive} component of the effective pair interaction between colloidal particles carrying 
charges of equal sign.  The same attraction is also invoked to explain the re-entrant liquid-solid 
coexistence described earlier \cite{59}.\\\\
However recent {\it direct} measurements of the effective interaction between pairs of colloidal particles 
show no evidence of an attraction, at least at low concentration.  Such direct measurements are based 
on video microscopy.  In the simplest method, a large number of instantaneous configurations of a 
dilute suspension are recorded, and the colloid-colloid pair distribution function $g(r)$ is calculated 
from the measured distribution of inter-particle distances \cite{66}.  The effective pair potential coincides 
with the potential of mean force in the low concentration limit \cite{8} and is hence given by :
\begin{eqnarray}
 v(r)=-k_{\rm B}T\log[g(r)]	 	
\label{23}
\end{eqnarray}
An alternative method follows the relative Brownian motion of a pair of colloidal particles released 
from initial positions where they were localized by optical tweezers \cite{67}.  Both sets of measurements 
confirm that the effective pair potential is purely repulsive, and may be fitted to a screened Coulomb 
(DLVO) form by adjusting the effective charge $Z^{*}$ and the inverse
screening length $\kappa_{\rm D}$,  compatible  
with estimated salt concentrations.  However this same DLVO potential, used within a harmonic 
approximation, seems to be unable to reproduce the measured bulk modulus of a colloidal crystal at 
much higher concentrations \cite{68a,68b}, although the difference between experiment and theory may be 
linked to the omission of the "volume" contribution to the effective interaction energy between 
polyions, to be discussed in the following section.
\subsection{DLVO theory revisited}
The effective pair potential between spherical polyions, first derived by Derjaguin and Landau, and 
Verwey and Overbeek \cite{58}, is easily recovered within the framework of DFT.  Upon tracing out the 
microion degrees of freedom, the effective interaction energy between polyions is given by eq. (\ref{8}), 
once the grand potential $\Omega$  has been determined from the variational principle (\ref{1}).   The required 
intrinsic free energy functional is defined by eqs. (\ref{5}) and
(\ref{7}).  Neglecting $F_{\rm corr}$ in the latter amounts   
to mean-field theory, and the Euler-Lagrange equation associated with the variational principle (\ref{1}) 
leads back to the PB equation (\ref{10}) for the local electrostatic
potential $\psi(\mathbf r)$ .  Since the external 
charge associated with $N$ polyions is a multi-centre distribution, the numerical task of solving the PB 
equation for $N$ interacting polyions is a formidable one, which can only be handled numerically 
using advanced optimization techniques, coupling Molecular Dynamics (MD) simulations with DFT 
\cite{69a,69b,70}.\\\\
A more phenomenological approach may be adopted, whereby the bare polyion charge is reduced to 
a considerably lower effective value by a Stern layer of tightly bound counterions.  The coupling 
between the polyions and the remaining microions (forming the so-called "diffuse" double-layer) is 
accordingly strongly reduced, so that the corresponding density
profiles $\rho_{\alpha}(\mathbf r)$  vary much more  
smoothly in the vicinity of the polyion surfaces.  In that case the integrands of the ideal 
contributions to F in eq. (\ref{7}) may be expanded to second order in
the deviations $\Delta \rho_{\alpha}({{\mathbf r}})=\rho_{\alpha}({\mathbf r})-n_{\alpha}$
  of the local densities from their bulk values, i.e. :
\begin{eqnarray}	 					
F_{\rm id}\approx\sum_{\alpha=\pm}\left\{F_{\rm
    id}(V,T,n_{\alpha})+\frac{k_{\rm
    B}T}{2n_{\alpha}}\int_{V}[\Delta\rho_{\alpha}(\mathbf
    r)]^{2}{\rm d}{\mathbf r}\right\}
\label{24}
\end{eqnarray}
where $F_{\rm id}(V,T,n_{\alpha})$  is the Helmholtz free energy of an
ideal gas of density $n_{\alpha}$ .  The free energy  
functional defined by (\ref{5}) and (\ref{6}), neglecting $F_{\rm
  corr}$, is now a {\it quadratic} functional of the local  
densities, and the resulting Euler-Lagrange equation reduces to a {\it
  linear} multi-centre PB (LPB)  
equation for the local potential $\psi(\mathbf r)$ , in the familiar
form : 
\begin{eqnarray}
\left(\nabla^{2}-\kappa_{\rm D}^{2}\right)\psi({\mathbf
r})=-\frac{4\pi}{\epsilon}\rho^{\rm ext}({\mathbf r}) 
\label{25}
\end{eqnarray}
This linear equation is easily solved by Fourier transformation in the bulk, with the boundary 
condition that the potential and its gradient vanish at infinity.
Since $\rho^{\rm ext}$  is a linear superposition of 
contributions from the $N$ polyions, the same is true of $\psi(\mathbf
r)$  and the resulting $\rho_{\alpha}(\mathbf r)$ .  Since microions 
cannot penetrate the spherical polyions, the excluded volume
condition, $\rho_{\alpha}({\mathbf r})=0;\;\;|{\mathbf r}-{\mathbf
R_{i}}|<a(1\le i\le N)$ 
 , must be imposed via a constraint, or by the use of a polyion-
microion pseudo potential \cite{40a,40b}.  Note that, strictly speaking, the linear superposition of densities 
only holds provided the electric double-layer associated with a given polyion does not overlap a 
neighbouring polyion ("weak overlap approximation").  This is not a significant limitation except at 
high polyion packing fraction.  The excluded volume condition leads to an additional polyion charge 
renormalization such that the effective charge $Z^{*}$ is multiplied
by the DLVO factor $e^{\kappa_{\rm D}}/(1+\kappa_{\rm D}a)$;  
the product will henceforth be designated by $Z^{*}$.  The local potential is finally of the form  :
\begin{eqnarray}
 \psi({\mathbf r})=\sum_{i=1}^{N}\psi^{(i)}({\mathbf
 r})=\sum_{i=1}^{N}\frac{Z^{*}e}{\epsilon}\frac{\exp\left({-\kappa_{\rm
 D}|{\mathbf r}-{\mathbf R_{i}}|}\right)}{|{\mathbf r}-{\mathbf R_{i}}|}	 	
\label{26}					
\end{eqnarray}
When (\ref{26}) and the resulting $\rho_\alpha({\bf r})$  are substituted into the quadratic free energy functional, the 
effective interaction energy (\ref{8})  between the $N$ polyions reduces to :
\begin{eqnarray}
V_{N}(\{{\mathbf R_{i}}\})=V_{0}+\sum_{i<j}v\left(|{\mathbf R_{i}}-{\mathbf
  R_{j}}|\right)
\label{27}					
\end{eqnarray}
where  $V_{0}(T, n, n_{+},n_{-})$ is a state-dependent "volume" term \cite{71}, the detailed expression of which is 
given in \cite{40b} and \cite{72}.  This term has been generally overlooked, but it strongly influences the 
calculated phase behaviour of charge-stabilized colloidal dispersions, as discussed later.  The main 
contribution to the volume term is the sum of cohesive electrostatic free energies arising from the 
electrostatic attraction between each polyion and its associated double layer of opposite charge.  The 
physical interpretation of $V_{0}$ is discussed in detail in ref. \cite{73}.  The effective pair potential is 
precisely of the well known DLVO form \cite{58}, namely:
\begin{eqnarray}
 v(R)=\frac{Z^{*2}e^{2}}{\epsilon}\frac{e^{-\kappa_{\rm D}R}}{R}	 			
\label{28}
\end{eqnarray}
Note that the pair-wise additivity in eq. (\ref{27}) is a direct consequence of the quadratic nature of the 
approximate free energy functional;  the exact ideal contribution (\ref{5}) to the free energy would lead to 
many-body effective interactions.\\\\
There are alternative contraction procedures to reduce the initial asymmetric, multicomponent PM to 
an effective one-component system, based on the OZ equations and approximate closures, like the 
MSA, for the partial pair distribution functions $g_{\alpha\beta}(r)$
\cite{10,11,12a,12b}.  The simple assumption of the 
asymptotic behaviour $c_{\alpha\beta}(r)\rightarrow
v_{\alpha\beta}(r)/k_{\rm B}T$  of the direct correlation functions is sufficient to show 
that the effective pair interaction is asymptotically of the DLVO form
(\ref{28}), and reduces to (\ref{28}) in the  
appropriate limits.  Corrections to the DLVO potential at short range may be obtained from 
numerical solutions of the fluid integral equations, and confirm the purely repulsive nature of the 
effective pair potential at low colloid concentration, at least when all microions are monovalent 
\cite{12b}.\\\\
The phase diagram of a monodisperse system of particles interacting via the DLVO pair potential 
(\ref{28}) has been determined by extensive MC simulations as a function of colloid and salt 
concentrations, the latter determining the inverse Debye screening
length $\kappa_{\rm D}$  \cite{38a,38b}.  As expected 
from the purely repulsive nature of the effective pair potential, the phase diagram exhibits a first-
order transition between a single (disordered) fluid phase and
(ordered) FCC or BCC crystalline phases, 
BCC being the stable phase in the $\kappa_{\rm D}\rightarrow 0$
limit, which corresponds to the widely studied "one-component plasma"
(OCP) model \cite{74}.  However the coarse-graining procedure leading from
the  
initial multicomponent polyion/microion "mixture" to the effective one-component system also 
introduces the volume term $V_{0}$  into the effective interaction energy between polyions.  Since $V_{0}$  is a 
{\it non-linear} function of the polyion density  $n$,  due  to  the
overall  charge  neutrality condition
$nZ^{*}+n_{+}z_{+}+n_{-}z_{-}=0$, this term must be included in the
free energy for a proper determination of the  
phase behaviour \cite{40a,40b,73}.  At high concentrations of added salt, the variation of $V_{0}$  with $n$ is 
sufficiently slow not to affect the phase behaviour significantly.  At
salt concentrations lower than $10^{-5}$M, the volume term drives a
van der Waals-like instability of the fluid, which separates into two  
phases of very different colloid concentrations, reminiscent of the "gas" and "liquid" phases of 
ordinary molecular fluids.  Depending on the effective colloid surface charge, this phase separation 
is completely disconnected from the freezing transition driven by the repulsive pair interaction (\ref{28}), 
and exhibits upper and lower critical points (re-entrant behaviour) in
the $n-n_{s}$ plane (see Figure 2);  or it merges 
with the freezing line, leading to a considerable broadening of the fluid-solid coexistence region, and 
the possibility of upper and lower triple points,as seen in Figure 3. The
possibility of a fluid miscibility gap in
polyelectrolytes had been conjectured already in 1938 by Langmuir, who referred to it as "unipolar 
coacervation" \cite{75}, and the complex phase scenario of references
\cite{40a,40b,73} has recently been 
confirmed by an extension of Debye-H\"uckel theory to the highly asymmetric PM \cite{76}.  These 
calculations provide a natural explanation for the observed phase behaviour of charge-stabilized 
colloids, including the formation of voids \cite{60,61,62,63,64,65},
{\it without} the assumption of a long-range  attractive 
component in the effective pair interaction which is frequently made \cite{77}, but lacks a firm theoretical 
basis.
\subsection{Beyond DLVO}
It was shown earlier that the effective force between two uniformly charged plates may turn 
attractive at short distances if correlations between microions are
properly accounted for.  A similar effect is expected to 
hold between two spherical polyions, as first conjectured for two isolated spheres by Patey \cite{14} on 
the basis of numerical solutions of the HNC equation.  When two spheres are sufficiently close, such 
that the shortest separation h between their surfaces is much less than their radius $a$, the problem 
reduces essentially to that of two charged planes discussed earlier :  a short-range correlation-
induced attraction may be expected for divalent counterions.  A collective polarization mechanism of 
adsorbed counterions operates in the opposite limit $h \gg a$, and leads to a "doubly screened" 
attractive component, proportional to $e^{-2\kappa_{\rm D}R}$ , which is always dominated by the DLVO 
repulsion \cite{78}.  It is important to stress that effective attractions between spherical polyions require a 
finite concentration of the latter in the absence of salt, i.e. when
only counterions are present \cite{78,79}.  No such restriction holds at finite salt concentration, where the problem of two isolated 
polyions is a meaningful limit.\\\\  
A similar "doubly screened" attraction between spherical polyions
occurs in the charge regulation PM introduced earlier
(c.f. eq. (12));  explicit HNC calculations have been carried
out for small mineral oxide particles \cite{26}.  Note that the
effective interaction between two isolated colloids is  
purely attractive (for $R > 2a$) in the case of a symmetrical  
adsorption of co and counterions ($V_{+} = V_{-}$ in eq. (12)),
because the colloidal particles are then, on average, neutral. 
\subsection{Computer simulations}
Density-functional molecular dynamics simulations were performed
for monovalent counterions without \cite{69b,81} and with \cite{80}
added salt. The effective interaction between the polyions was   
found to be repulsive, and in quantitative agreement with DLVO theory,
provided the charge was renormalized according to the cell model
prescription incorporating the  excluded volume correction \cite{Irene2}.
This approach includes  counterion correlations, but is not completely equivalent to a  
full simulation of the PM since an approximate polyion-counterion 
pseudopotential is used and the exact density functional of the
electrolyte is unknown. Due to the high charge asymmetry between poly-
and microions, 
 a full simulation of the PM requires the inclusion of many counter-
 and salt ions and is not feasible on present-day computers. There are
 two ways to escape from this: 
either by considering only a small number of polyions or by
reducing the charge asymmetry significantly.
By considering only two polyions within the PM, the effective pair
potential can be calculated within a simulation by averaging over the
microscopic ions, while keeping the polyion positions fixed. For
monovalent microions, the   
simulation yields repulsive forces correctly described by DLVO theory  
 both with \cite{Irene1} and without \cite{83}
added salt. For divalent counterions, attractive forces are obtained,  
which may be  attributed to counterion correlations  \cite{79} and
Coulomb depletion \cite{82}, resulting from  
a depletion zone of counterions between two nearly  
touching polyions due to the strong counterion repulsion.
 A typical snapshot of the counterions
around two fixed polyions and the averaged force versus polyion distance
are shown in Figure 4. A system of three polyions was simulated in order  
to extract effective  triplet interactions which were found, however, to be  
small with respect to the pairwise part \cite{84}.\\\\ 
On the other hand, there are quite a number of full simulations of the
PM with a reduced charge asymmetry corresponding to micellar rather
than colloidal polyions. Most of the work is summarized in a  review
by Vlachy \cite{Vlachy}. 
As for more recent work, the role of salt valency has been investigated
in some detail \cite{86}. Moreover  Linse and Lobaskin pushed
the  charge asymmetry to $Z:z_+ = 60:1,\;60:2,\;60:3$ using efficient
cluster move algorithms \cite{13,Linse} and treating 80
polyions simultaneously \cite{85}. For the $60:3$ asymmetry
clear indications for a phase separation were 
reported which can be attributed to an effective attraction between
polyions. 
\section{Polyions in confined geometries}
The bulk behaviour of polyionic dispersions is expected to change significantly in the presence of 
neutral or charged interfaces confining the dispersion to a restricted volume.  Such confinement 
occurs naturally in the vicinity of the sample container, which may be approximated locally by an 
infinite plane restricting the suspension to a half space.  Many experiments are carried out in a slit 
geometry, with the suspension confined between parallel planes (e.g. glass plates);   if the interplanar 
spacing is comparable to the diameter $2a$ of the polyions, the suspension behaves like a quasi-two-
dimensional (2d) many-body system exhibiting interesting phase behaviour \cite{87}. 
\\\\
The electrostatic effects of confinement are essentially threefold. (i)  The interface between two 
different dielectric media, e.g. the suspension and the glass of the container wall, must be 
characterized by appropriate boundary conditions;  in particular the dielectric discontinuity implies 
the presence of electrostatic image charges. (ii).  The confining surfaces lead to a reduction in 
screening power of the electrolyte, which ceases to be exponential in directions parallel to the 
surface \cite{88}.  (iii)  If the surfaces are charged, they attract or repel the polyions and microions, and 
release additional counterions into the suspension.
\\\\
These effects will modify the electrostatic interactions between double-layers associated with 
polyions close to the confining walls.  This has been clearly demonstrated by a number of direct 
measurements of the effective forces between spherical polyions confined near a glass wall, or in 
narrow slits, using digital video microscopy techniques similar to those mentioned earlier for 
measurements in the bulk.  Measurements of the colloid pair distribution function $g(r)$ carried out at 
sufficiently low concentration allow the effective pair potential to be extracted directly from eq. (23) 
\cite{89,90}, while data collected from more concentrated samples, where the effective potential $v(r)$ is 
expected to differ significantly from the potential of mean force defined by eq. (\ref{23}), require an 
elaborate inversion procedure based on fluid integral equations (properly adapted to a 2d geometry) 
\cite{91,92}, or on an iterative method involving computer simulations \cite{89}.  These experiments leave 
little doubt for the existence of an {\it attractive} component of the effective polyion pair potential  at  
{\it low}  ionic strength  (i.e. for  salt concentrations estimated to be below about $10^{-5}$M), when the 
polyions are highly confined, as achieved in narrow slits \cite{89,91}, or by localizing them close to a 
single plane by optical tweezers \cite{90,93,94}.  The observed attractive well in $v(r)$ is relatively 
shallow (depth of the order of $k_{\rm B}T$), and long-ranged;  typically $v(r)$ becomes negative for 
interparticle distances considerably larger than the particle diameter $2a$.  The experiments carried out 
for two particles close to a charged plate with the centre-to-centre vector parallel to the plate, clearly 
show that the attractive well disappears when the particles are moved away from the plate;  the 
measured $v(r)$ returns to its bulk DLVO form (\ref{28}) when the distance to the plate is several particle 
diameters \cite{90,93,94}.    An additional twist for suspensions of colloidal particles confined to a slit is 
provided by the introduction of quenched obstacles, in the form of larger, charged colloids, which 
form a disordered porous 2d matrix.  As the concentration of such obstacles is increased, the 
effective interaction potential between the smaller colloids is
observed to develop a second attractive 
component extending to larger inter-particle separations \cite{95}.  Closely related observations of 
metastable colloidal crystallites suggest that facets of such crystals behave very much like charged 
plates, inducing effective attractions between nearby polyions \cite{93} due to many-body interactions.
\\\\
The DFT formulation of DLVO theory in the bulk has been extended to the case of confined 
particles.  The electrostatic potential due to a point polyion near a planar surface separating two 
media of dielectric constants $\epsilon$  and $\epsilon^{'}$  was determined within LPB theory by Stillinger as early as 
1961 \cite{96}.  The case of two spherical polyions of finite radius a near an uncharged planar surface 
was examined within LPB theory in \cite{97}.  Since microions cannot leak beyond the dividing surface, 
the spherical symmetry of the electric double layers around each of the polyions is broken.  The 
effective repulsion between the polyions is enhanced, and decays like $1/R^{3}$, where $R$ is the separation 
parallel to the surface, as expected from the considerations in \cite{88}. 
\\\\
If the confining surface carries an electric charge density $\sigma$ , as would be the case for a glass plate in 
contact with an aqueous dispersion, the planar electric double-layer building up near the surface will 
modify the local co and counterion concentrations in the vicinity of nearby polyions, and hence 
influence their mutual interaction.  Considering the charged wall as an additional colloidal particle of 
infinite radius, it is clear from the inherent pair-wise additivity that the quadratic free energy 
functional, defined by eqs (\ref{24}) and (\ref{7}) (with $F_{\rm
  corr}=0$), which underlies LPB theory, will not affect  
the effective interaction between two polyions near the wall.  Clearly the intrinsic three-body nature 
of the problem must be taken into account.  The minimal theory to achieve this is to carry the 
expansion (\ref{24}) one order further, allowing for a direct coupling between the three double-layers 
associated with the polyions and surface.  Application of DFT perturbation theory \cite{69b} leads to the 
prediction of a long-range attraction between polyions, decaying like $1/R^{3}$  along the surface, and 
exponentially normal to the surface, for finite concentrations of polyions, to allow for an imbalance 
between co and counterion concentrations \cite{98}.  Both the location and depth (typically of the order 
of $k_{\rm B}T$) of the predicted potential well are in semi-quantitative agreement with direct experimental 
measurements \cite{90,93,94}.  The attraction predicted within mean-field theory occurs when the 
surfaces of the spherical polyions are several Debye lengths apart, and has nothing to do with the 
correlation-induced-attraction at very short range predicted in the bulk, and which is significant only 
for multivalent counterion \cite{78,79,82,85,86}.\\\\ 
The generalization of DLVO theory may also be extended to slit \cite{99,100} and pore \cite{101} 
geometries.    The quasi-1d confinement in the latter case leads to an even slower decay of the long-
range attraction.  The existence of an effective attraction between two like-charged polyions 
confined by a charged cylindrical surface was also predicted on the basis of numerical solutions of 
the non-linear PB equation \cite{102}, but it has since been proved rigorously that PB theory can only 
lead to purely repulsive effective interaction between two identical polyions confined to a cylinder of 
arbitrary cross-section \cite{103,104}, whatever the boundary conditions at the charged surfaces.  This 
proof does not invalidate the results of the perturbation theory \cite{98,100,101}, since charged 
renormalization (which accounts for short-range correlations), followed by an LPB perturbation 
treatment of the non-adsorbed counterions is not a mere approximation to non-linear PB theory, 
which is, moreover, unphysical near highly charged surfaces.\\\\
A direct simulation of the PM for confined colloids
is only feasible for low surface charges and small 
particle sizes \cite{105}. In this
regime,  the attraction between colloids is strongly
enhanced by the presence of a charged wall with respect to that 
in the bulk \cite{82}. Also the wall-particle 
interaction was found \cite{105}
to be attractive for divalent counterions.
\section{Rod-like polyions}
There is a large variety of rod-like polyions 
ranging from  synthetic suspensions to biological macromolecules.
These include colloidal $\beta$-FeOOH \cite{ref7}, imogolite
\cite{Donkai}, boehmite \cite{ref8,Buining,Wierenga}, 
polytetrafluoroethylene \cite{ref9}, 
ellipsoidal polystyrene latex particles \cite{ref10},
cylindrical micellar aggregates \cite{Hendrikx,Wu} as well as
virus solutions from tobacco-mosaic (TMV) \cite{Bawden,Fradenreview}, 
or bacterial fd \cite{Graf,Dogic} and Pf1 viruses \cite{ref4}.
Another motivation to study rod-like colloids comes from the 
self-assembly of charged stiff biopolymers such as 
DNA strands \cite{DNAreview1,DNAreview2}, F-actin fibers 
\cite{actin} and  microtubules \cite{actin}
which constitute  rod-like polyions on a
supramolecular rather than on a colloidal length scale.\\\\
The simplest model system is the electric double layer around 
a single infinitely long cylindrical polyion 
of radius $a$ which is homogeneously  charged with line charge density
$\lambda$. LPB theory leads to the electric potential
$\Psi (r) = \lambda K_0(\kappa_{\rm D} r )/ a K_1(\kappa_{\rm D} a) \epsilon $ 
where $r$ is the distance from the rod axis
and $K_0(x),K_1(x)$ are  Bessel functions of imaginary argument.
Consequently, LPB theory predicts the  effective interaction $v(r)$ between two parallel rods per unit length 
to be of the following form:
\begin{eqnarray}
 v(r) =  \frac{\lambda^2}{(\kappa_{\rm D} a K_1(\kappa_{\rm D}
 a))^2\epsilon} K_0(\kappa_{\rm D} r ) 
\label{28b}
\end{eqnarray}
 where the factor $1/[\kappa_{\rm D} a K_1(\kappa_{\rm D} a)]$ is the excluded volume
 correction. Contrarily to the three-dimension
al case,   PB theory can
 be solved analytically within a cylindrical cell around the charged
 rod for vanishing salt concentration 
\cite{Morawetz,17}. The analytical 
solution for the counterion density field around
the colloidal rod exhibits a strong peak at contact \cite{Barratreview}
providing a theoretical framework of Manning
counterion condensation \cite{Manning,Holm}. As for spherical
macroions, one cannot extract an effective interaction from the PB
solution for  
a single double-layer directly. One possibility is to proceed  as for  the spherical 
PB cell model. Matching the LPB theory at the cell boundary \cite{HL94_2},
one derives the   effective interaction potential 
$v(r) =  {{{\lambda}^{*2}}/ \epsilon} K_0(\kappa r )$, 
where $\kappa$  and ${\lambda}^*$
are renormalized according to the cell model prescription.\\\\ 
A more direct theory, which requires a larger numerical effort, is to
solve the two-dimensional PB theory  for  two parallel rods in a slit 
geometry and extract the effective
force from there. In a recent numerical
study \cite{Ospeck} no attraction was found, consistent with the exact
result  \cite{103,104} in three dimensions.\\\\ 
Beyond the PB level of description, 
computer simulations for the effective interaction between
two parallel, homogeneously charged rods have been performed 
 for divalent counterions and no added salt \cite{GJensen}.
An effective attraction between the rods was found. The
question is  whether the attraction is due 
to correlations \cite{Comment} or  to fluctuations
\cite{Reply,Liu1}.
If the former is true, the attraction should increase with decreasing
temperature and may even persist for zero temperature, while fluctuation-
induced attraction should increase with temperature.
On the basis of a simple model \cite{Stilck2}, Levin and coworkers 
gave significant support
for a correlation picture. This can be intuitively
understood as the gain in electrostatic energy upon bringing together
staggered arrays of adsorbed counterions
\cite{53,Shklovskii}. 
Fluctuation \cite{Ha,Ha2} and
polarization effects \cite{Solis} may still play an important
additional  role.
For parallel rods on a triangular lattice, a negative pressure was
 found by computer simulation with  divalent counterions \cite{Tang},
 which can be attributed to  some attractive component in the
 effective interactions. \\\\
Qualitatively, the situation closely resembles
 the case of spherical macroions. But there are also differences.
First, Manning condensation theory applies only to rod-like polyions.
A possible mechanism of attraction of nearly touching 
rods by sharing Manning-condensed
counterions was proposed, which is only possible for rods \cite{Ray}.
Secondly, the Coulomb depletion mechanism for attraction \cite{82}
  was  found to be irrelevant for parallel rods.\\\\
For finite rod lengths and arbitray orientations of the rods, the
effective interaction  between two rods will depend both on the rod orientations
and on their center-of-mass separation. Starting from 
point charges distributed along the rods, LPB theory 
results in a Yukawa-segment model \cite{Klein1} which
 was confirmed by density-functional Molecular-Dynamics  simulations 
 for monovalent counterions \cite{HL94_2}, provided the charge and the
screening constant $\kappa$ are renormalized according
to the cell model prescription. The associated phase diagram for
 parameters suitable 
for the TMV suspension was calculated, involving different liquid
crystalline phases \cite{Graf2}.\\\\
A full simulation of the PM was performed for 
stiff polyelectrolytes \cite{Stevens99}, resulting in bundle
formation, which is a 
possible  sign of an effective  attraction. Clearly,
the effective pair potential
picture is insufficient for bundles \cite{Podgornik}
and many-body interactions play a significant role \cite{Ha,Ha2}.
\section{Lamellar polyions}
Lamellar polyions may be schematically looked upon as rigid or flexible
charged surfaces or platelets, providing a 2d counterpart of charged rods or
polyelectrolytes.  Examples are the geologically and technologically
important smectite clays \cite{106} and 
self-assembled bilayers of ionic surfactants, which constitute the
proto-type of biological 
membranes \cite{45}.\\\\
Consideration will first be given to electric double-layers around
infinitely thin, uniformly
charged circular or square platelets, which constitute a reasonable model
of smectite clay
particles, and in particular of the widely studied synthetic Laponite
mineral particles.  To a
good approximation the latter are rigid, thin discs, of
thickness $d\approx 1$nm, radius $a\approx 15$nm, carrying a structural
surface charge $\sigma\approx -e/(\rm nm)^{2}$.  Natural montmorrilonite 
clays are
silicate mineral platelets of similar chemical composition and crystal
structure, but of
irregular and polydisperse shape, and of much larger lateral dimension,
implying some
degree of bending flexibility.\\\\
Dry powders of clay will swell upon the addition of water, releasing
counterions into the
interlamellar volume, and building up interacting electric double-layers;
the swelling is
essentially driven by the double-layer repulsion of mostly entropic origin.
During the
initial stages of swelling, the spacing $h$ between platelets remains small
compared to their
lateral extension, so that a moderately swollen lamellar phase may be
reasonably modelled
by a stack of infinite charged planes \cite{107}.  The results for planar
geometry, discussed
earlier, apply then directly to the swollen phase.  In particular, limited
swelling, often
observed in the presence of divalent counterions, may be related to the
cohesive behaviour
(effective attraction between planes, or negative disjoining pressure) due
to correlation
effects within the PM \cite{108}.  An important issue in the understanding of
the swelling
behaviour is the competitive condensation of counterions of different
valence and/or size
\cite{108,109}.  The PM model is, however, expected to be inadequate when the
interlamellar
spacing h is only of the order of a few microion diameters.  Under those
conditions the
molecular nature of the solvent (water) must be explicitly taken into
account.  Constant
pressure MC simulations clearly show the importance of counterion hydration  
in
determining the swelling behaviour, and point to the role of $\rm K^{+}$ ions as a
clay swelling
inhibitor \cite{110a}.  When swelling proceeds until the interlamellar spacing $h$
becomes
comparable to the platelet radius $a$, finite size (edge) effects become
important.  The force
acting on a platelet follows from the integration of the stress tensor
$\underline{\underline\Pi}$
over the two faces $\Sigma_{+}$ and $\Sigma_{-}$ of the platelet:
\begin{eqnarray}
{\mathbf F}=-\int_{\Sigma_{+},\Sigma_{-}}\underline{\underline\Pi} {\rm d} \mathbf s 
\label{29}
\end{eqnarray}
where  ${\rm d} {\mathbf s}={\mathbf {\hat n}  }{\rm d}s$ is the
surface element oriented along the outward normal ${\mathbf {\hat n}
  }$, and the stress tensor has the standard form :
\begin{eqnarray}
\underline{\underline\Pi}=\left[P({\mathbf r})+\frac{\epsilon}{8\pi}|{\mathbf
  E}|^{2}\right]{{\mathbf I}}-\frac{\epsilon}{4\pi}{\mathbf E}({\mathbf
  r})\otimes{\mathbf E}({\mathbf r})
\label{30}
\end{eqnarray}
which generalizes the uniaxial expression (\ref{20}).  Within LPB theory, eq.
(\ref{29}) leads to the
following expression for the force between two coaxial discs separated by
$h$, immersed in
an ionic solution of inverse Debye length $\kappa_{\rm D}$  \cite{110b}:
\begin{eqnarray}
F_{z}(h)=(\pi
a^{2})\times\frac{4\pi\sigma^{2}}{\epsilon}\int_{0}^{\infty}J_{1}^{2}(x)\frac{1}{x}\exp\left\{
-\frac{h}{a}\sqrt{x^{2}+\kappa_{D}^{2}a^{2}}\right\}{\rm d}x 	     
\label{31}                                				
\end{eqnarray}
Numerical solutions of the non-linear PB equation, for the same coaxial
geometry, show
that the LPB expression (\ref{31}) strongly overestimates the force \cite{111}, as in
the spherical case,
thus requiring a proper renormalization of the surface charge density $\sigma$.
Recent MC
simulations show that the force may become attractive for divalent
counterions, as in the
case of infinite charged planes, but that the finite size of the discs
leads to significant
differences in the relative weight of electrostatic and contact
contributions \cite{112}.\\\\  
Finite concentration effects in highly swollen stacks of parallel platelets
of finite size may be
examined within a cell model, compatible with the shape of the platelet.  A
coaxial
cylindrical cell is chosen for a disc-shaped platelet, while a
parallelepipedic cell is better
adapted to square platelets \cite{110b,113}.  Note that the platelet
concentration determines the
cell volume, but not the aspect ratio of the cell, e.g. the ratio $R/h$ in
the case of a cylindrical
cell of radius $R$ and height $h$ equal to the inter-lamellar spacing;  the
optimum aspect ratio
for given volume and electrolyte concentration is determined by minimizing
the free
energy of the microions in the cell.  The latter has been calculated within
LPB theory which
can be solved analytically \cite{110b,113}, and within non-linear PB theory,
which requires
numerical solution;  the latter is greatly simplified within a Green's
function formulation
\cite{111,114}.\\\\
The osmotic pressures calculated as a function of Laponite concentration
within PB theory
agree reasonably well with experimental data \cite{115}, for reservoir salt
concentrations of the
order of $10^{-3}$M or larger, but differ dramatically from the predictions of
1d PB theory for
stacks of infinite platelets, pointing to the importance of edge effects in
highly swollen clays
\cite{114}.\\\\
In very dilute dispersions of clay platelets, such that the distance
between the centres of
neighbouring polyions is significantly larger than the particle radius, the
platelets can rotate
more or less freely (sol phase).  As the concentration increases, a gel
phase is formed,
depending on salt concentration.  Several recent experiments have attempted
to establish a
link between the mesoscopic fractal structure of the gel, and its
rheological properties \cite{115,116,117}, but no clear-cut scenario has yet emerged, due to metastability
and ageing of the
dispersions \cite{118}.\\\\
A theoretical description of the sol-gel transition of clay dispersions
hinges on a knowledge
of the highly anisotropic effective interaction between charged platelets
in an electrolyte.
The charged segment, or site-site model, introduced earlier for charged rods
has been
generalized to charged circular platelets, and used in MD simulations of
Laponite
dispersions \cite{119}.\\\\
For well separated platelets, the effective pair potential reduces to a sum
of {\it screened} interactions between the electrostatic multipoles
associated with the anisotropic electric
double layers around each platelet \cite{120}.  A simplified version of such an
effective
interaction, involving infinitely thin discs carrying an unscreened
quadrupole moment, has
been used in MC simulations which predict a reversible sol-gel transition
\cite{121}.\\\\
Lipid bilayers and membranes constitute another class of lamellar polyions,
which are
flexible.  Electrostatic interactions renormalize the bending rigidity of
these flexible
membranes \cite{45}, a subject of ongoing work beyond the scope of this review.
In relation to
the correlation-induced attraction between like-charged planes discussed
earlier, the fluidity
of membranes provides an additional mechanism for attraction between
membranes,
resulting from the lateral charge fluctuations within the planes of the
latter \cite{122}.
\section{Discrete solvent effects}
So far all theoretical considerations of interacting electric double-layers
were based on the
PM, which ignores the molecular nature of the solvent.  The PM thus
neglects excluded
volume effects of the latter, as well as hydration of ions and the expected
reduction of the
local dielectric constant near highly charged surfaces, due to polarization
of nearby water
molecules.  Neglect of these and other solvent effects is expected to be
particularly
inadequate in situations where the spacing between two charged polyion
surfaces is only of
the order of a few molecular diameters.

To go beyond the PM level of description, simple models have been used to
describe the
solvent molecules (generally water).  The crudest model is to represent the
latter by neutral
hard spheres of appropriate diameter, while keeping a macroscopic
dielectric constant $\epsilon$  in
the Coulombic interactions between ions;  this model accounts only for
excluded volume
effects.  The next refinement is to consider hard spheres with embedded
point dipoles, to
account for the highly polar nature of the solvent \cite{8,123}, while
a reasonable local coordination of the solvent molecules can only be achieved by adding higher order
multipoles \cite{124,125,126}.  In MC or MD simulations, much more sophisticated
pair potentials
may be used, involving three or more interaction sites on each water
molecule, as in the
widely used SPC/E potential \cite{127}.\\\\
A purely HS solvent is implicit in the modified PB formulation of
Kralj-Iglic and Iglic \cite{128}
and others \cite{101,129}.  As expected, the modified PB equation leads to a
saturation of the
counterion density at contact, for high surface charge densities.  The
force between two
charged surfaces in a mixture of charged and neutral hard spheres was
calculated by Tang
et al. \cite{130} within a DFT generalizing the earlier theory of the same
authors for a PM
electrolyte \cite{52};  the force is a strongly oscillatory function of the
spacing between the
plates, due to HS layering, and is rather insensitive to surface charge and
salt concentration.
The latter observation, which holds when solvent, anions and cations are of
the same size,
does not carry over to the more realistic case of unequal diameters, which
has been
investigated within IHNC theory \cite{131}.  At low surface charge, the total
force between
platelets is reasonably approximated by a superposition of the pure HS
solvent "hydration" force, and the electrostatic contribution of the ions, as
calculated within the
PM, in the case of equal diameters \cite{131,132}.
\\\\
The dipolar HS model for the solvent was first used to determine the
structure of the
electric double layer near a single charged wall within the MSA \cite{133,134}.
More accurate
and complete results on the planar electric double layer in a dipolar
solvent were obtained
from numerical solutions of the coupled RHNC equations for the three
density profiles
\cite{136}.  These calculations give Statistical Mechanics evidence for the
reduction of the local
dielectric constant   near the charged surface, and for significant
electrostriction.  A generic
density functional, based on Rosenfeld's successful "fundamental measure"
theory for hard
core fluids \cite{137,138} has been put forward, which can be adapted to any
polyion geometry
\cite{139};  when applied to a single planar double layer, this theory shows a
considerable
enhancement of the counterion density at contact, when a dipolar, rather
than bare HS
solvent is used.\\\\
The effective solvation force between two charged plates immersed in an
ionic solution of
charged and dipolar HS was calculated within a quadratic free energy
functional of the local
density, charge density and polarization, generalizing eqs. (\ref{24}) and (\ref{6})
(with direct
correlation functions approximated by bulk MSA solutions) \cite{new_ref}.  This
calculation provided
the first convincing evidence of the strong influence of a granular (as
opposed to
continuous) solvent on the solvation forces at short range.
\\\\
The most complete investigation so far of the potential of mean force
between two
spherical polyions immersed in an ionic solution with a "realistic" solvent
involving hard
spheres with point dipoles and tetrahedral quadrupoles \cite{125}, appears to be
the work by
Kinoshita et al \cite{140}, who solved the RHNC equations for this highly
asymmetric
multicomponent "mixture", for ratios $10\le a/a_{s}\le 30$, where $a$ and
$a_s$ are the polyion and solvent molecule radii.  While maintaining $a_{s}$
fixed at 0.14nm, a value 
appropriate for water,
these authors varied the counterion radius.  Their most striking finding is
that larger
counterions are more strongly adsorbed to the polyion in the presence of a
molecular
solvent, leading to a greater reduction of the Coulomb repulsion between
the polyions,
and even to the possibility of an effective attraction for monovalent
counterions, when
their radius exceeds $a_{s}$ by more than $20\%$.  The trend towards a reduction of
effective
repulsion with increasing counterion size is the exact opposite of the
prediction of the PM, a
clear illustration of the pitfalls of the latter in describing the
short-range behaviour of
solvation forces!  Obviously more work on discrete solvent models is needed.
\section{Outlook}
Although the understanding of 
 effective interactions between electric double layers
has clearly advanced in the last decade through a combination of
 new theoretical approaches,
quantitative measurements and large-scale computer simulations,
there are still many open questions. Future research should
focus on the influence of image charges \cite{Tandon} due to dielectric
discontinuities between the solvent and the container
walls or the polyions themselves. 
Real samples possess, moreover, an intrinsic polydispersity
in size, charge and shape which becomes relevant for
a quantitative comparison between theory and experimental data.
Another rapidly growing field concerns flexible polyelectrolyte chains
whose stiffness is governed by their persistence length 
\cite{Barratreview}. A theoretical approach needs
input from both nonlinear screening  and polymer theories.
Colloids can also be stabilized by polyelectrolytes, and the resulting
effective interaction becomes qualitatively  different from that
between charged spheres, particularly near the overlap concentration.
The microscopic incorporation of the solvent is still in its early stage 
and full molecular theories describing hydration forces and hydrogen bonding
are highly desirable for aqueous suspensions. Lastly alternative 
approaches, such as recent field theoretical formulations
\cite{Netz}, might lead to additional insight.
\section*{Acknowledgements}
We thank E.Allahyarov, D.Goulding, Y. Levin, P. Linse, V. Lobaskin,
P. Pincus, and R. van Roij for helpful comments.

\begin{center}
\begin{figure}
\epsfig{file=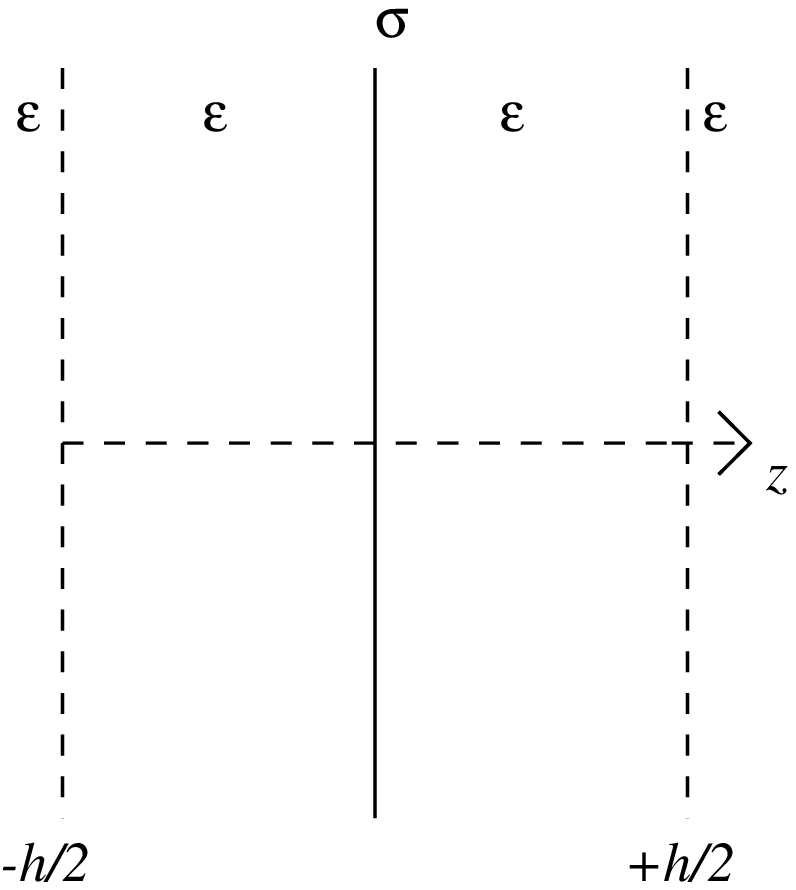,width=8cm}
\label{fig1a} 
\end{figure}  
\end{center}
\begin{center}
\begin{figure} 
\epsfig{file=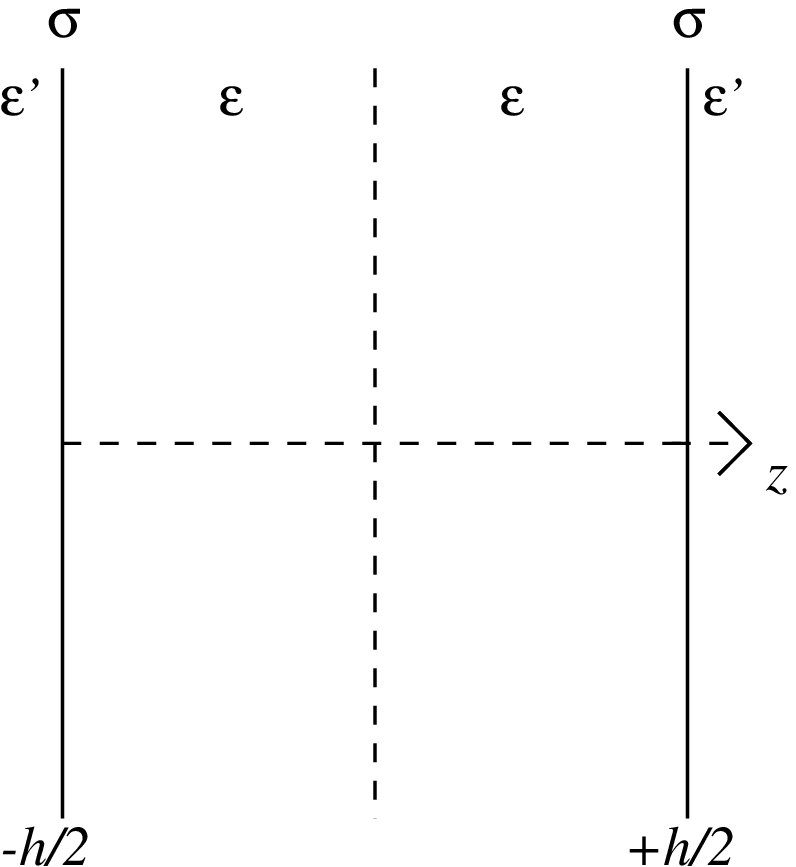,width=8cm}
\label{fig1b} 
\caption{a,b}
\end{figure}  
\end{center}  
\begin{figure}
\epsfig{file=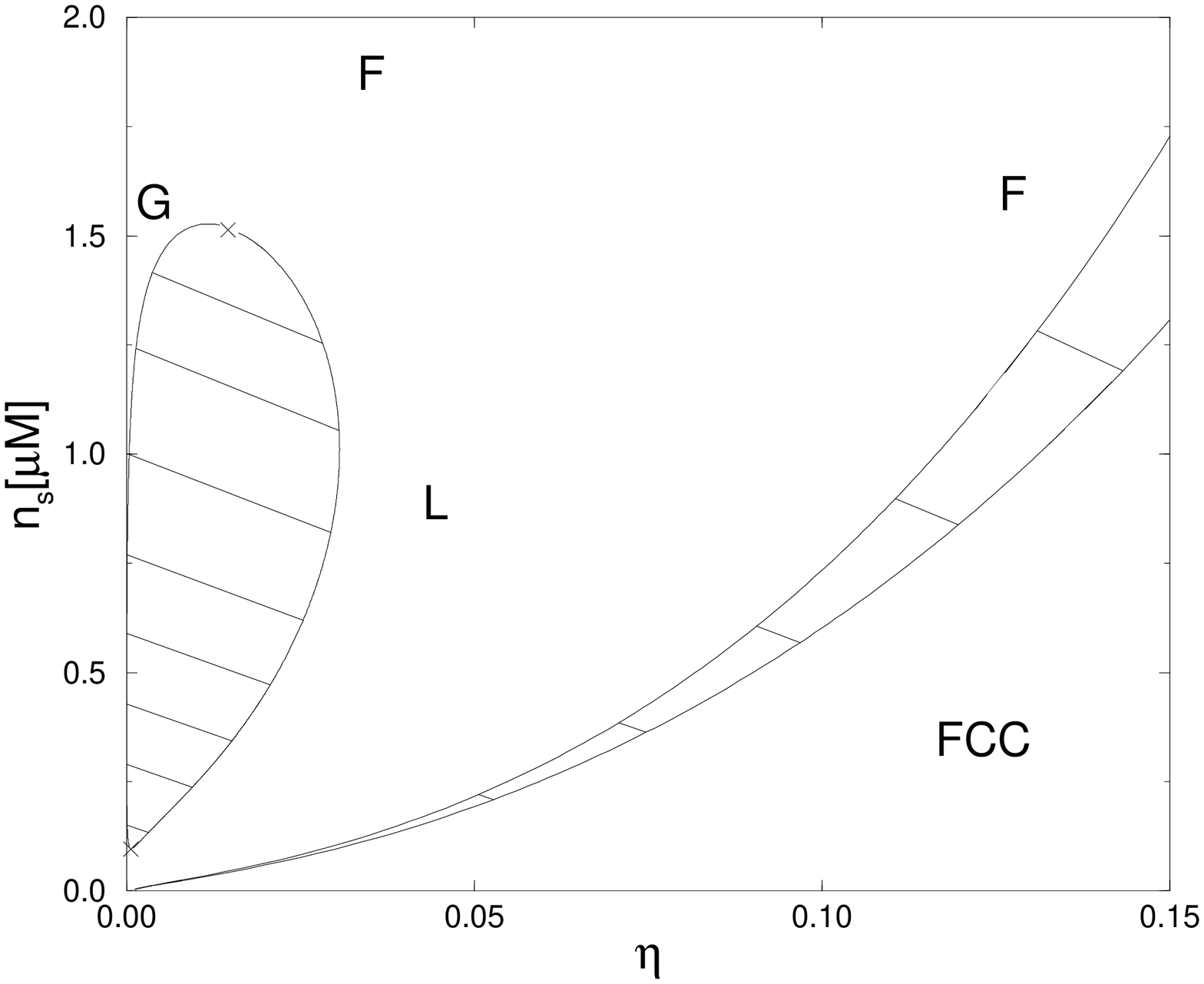,width=13cm,angle=-90}
\label{fig2}
\caption{ } 
\end{figure}
\begin{figure}
\epsfig{file=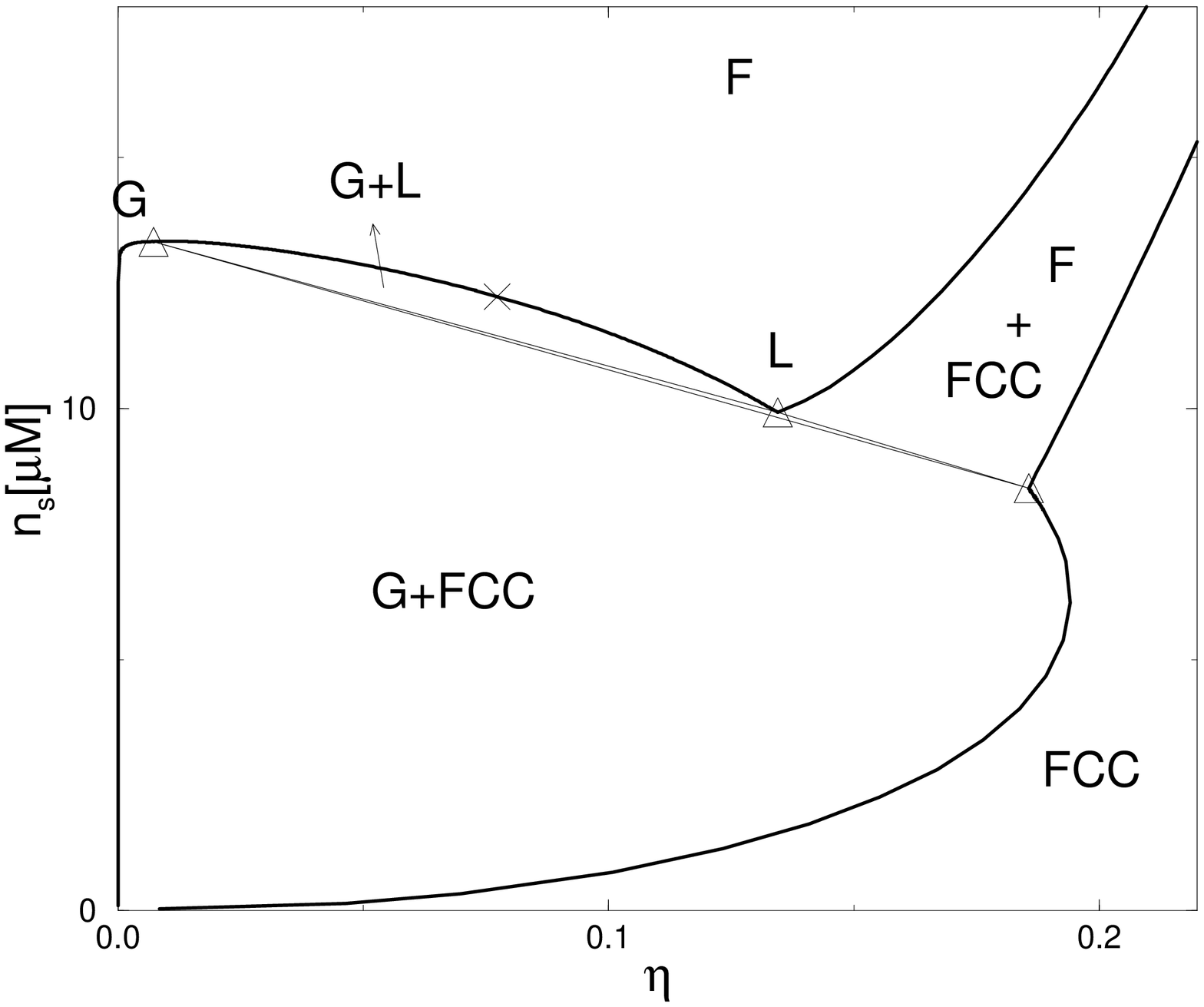,width=13cm,angle=-90}
\label{fig3}
\caption{ }
\end{figure} 
\begin{figure}
\epsfig{file=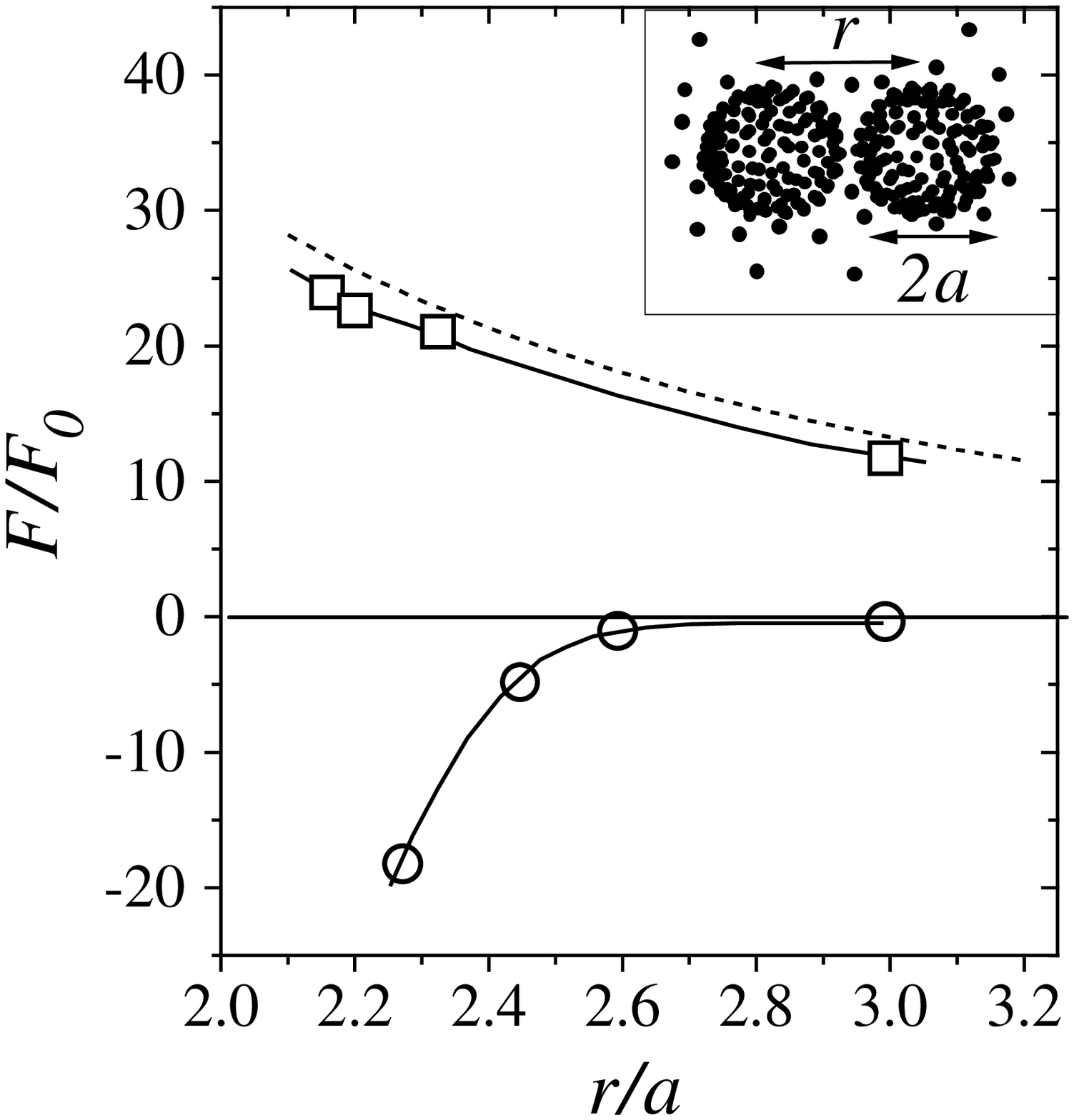,width=15cm}
\label{fig4} 
\caption{ }
\end{figure} 
\newpage
\begin{center}
{\large  Figure captions}
\end{center}
Figures 1a, 1b: Two different plate geometries.\\\\
Figure 2: Phase diagram for a charged suspensions
in the $\eta - n_{s}$ plane with $\eta=4\pi n a^3/3$ and $Z=350$, $a=380$nm
, $\epsilon=13.9$ (penthanol) and monovalent counterions at room
temperature. Possible phases are fluid (F), gas (G), liquid (L), and
face-centered-cubic (FCC) crystals. Tie-lines denote coexistence
conditions.\\\\
Figure 3: Same as Figure 2, but now for $Z=1000$,
$a=350$nm, $\epsilon=25.3$ (ethanol). There is a triple point of
coexisting gas, liquid and crystal.\\\\
Figure 4: Effective force $F$ in units of
$F_{0}=\frac{Z^{2}e^{2}}{a^{2}}\times 10^{-4}$ between two
polyions versus reduced distance $r/a$. The squares are for aqueous
suspensions. The force is repulsive in good agreement with DLVO-theory
(dashed line). The circles are for a solvent with a strongly reduced
dielectric constant $\epsilon$ and show attraction. The parameters are
given in detail in Ref.\ \cite{82}. The inset shows a counterion
configuration around two fixed polyions.

\end{document}